\documentclass[a4paper,11pt]{article}

\usepackage{jheppub} 
\usepackage[T1]{fontenc} 
\usepackage{slashed}
\usepackage[utf8x]{inputenc}
\usepackage{color}
\usepackage[normalem]{ulem}
\usepackage{bm}

\allowdisplaybreaks

\title{An effective field theory for non-relativistic Majorana neutrinos}

\author{S. Biondini,}
\author{N. Brambilla,}
\author{M. A. Escobedo}
\author{and A. Vairo}

\affiliation{Physik-Department, Technische Universit\"{a}t M\"{u}nchen, \\James-Franck-Str. 1, 85748 Garching, Germany}
\preprint{TUM-EFT 34/12}

\emailAdd{simone.biondini@tum.de}
\emailAdd{nora.brambilla@ph.tum.de}
\emailAdd{miguel.escobedo@ph.tum.de}
\emailAdd{antonio.vairo@ph.tum.de}

\abstract{Heavy Majorana neutrinos enter in many scenarios of physics
  beyond the Standard Model: in the original seesaw mechanism they provide a
  natural explanation for the small masses of the Standard Model
  neutrinos and in the simplest leptogenesis framework they are at the origin of
  the baryonic matter of the universe. In this paper, we develop an
  effective field theory for non-relativistic Majorana particles,
  which is analogous to the heavy-quark effective theory. Then, we
  apply it to the case of a heavy Majorana neutrino decaying in a hot
  and dense plasma of Standard Model particles, whose temperature is
  much smaller than the mass of the Majorana neutrino but still much larger 
  than the electroweak scale. 
  The neutrino width gets zero-temperature contributions that 
  can be computed from in-vacuum matrix elements, and thermal corrections.
  Only the latter will be addressed.
  Symmetry and power counting arguments made manifest by the effective 
  field theory restrict the form of the thermal corrections and simplify their calculation.
  The final result agrees with recent determinations obtained with different methods.
  The effective field theory presented here is suitable to be used 
  for a variety of different models involving non-relativistic Majorana fermions.}

\begin{document} 
\maketitle
\flushbottom

\section{Motivation and introduction}
Neutrino flavour oscillations, the large matter-antimatter asymmetry
of the universe and dark matter are commonly interpreted as major
experimental observations that require going beyond the Standard Model
(SM) of particle physics.  Among the many possible extensions of the
SM that have been proposed, a minimal extension would consist in the
inclusion of some generations of right-handed neutrinos.  Right-handed
neutrinos are singlet under the SM gauge groups, therefore they are
often called sterile neutrinos.  Models have been considered with
different sterile neutrino generations and with neutrino masses
spanning from the eV scale to $10^9$ GeV. As we are going to list
briefly in the following paragraph, these models provide some natural
explanations of the above observations.  We refer to~\cite{Drewes:2013gca} 
for a recent review and a large body of references.

The experimental observation of neutrino mixing~\cite{Fukuda:1998mi,Ahmed:2003kj} 
implies that neutrinos carry a finite mass. 
A simple model capable of giving mass to the observed SM neutrinos and
at the same time providing a natural explanation for its smallness is
the seesaw mechanism originally proposed in~\cite{Minkowski:1977sc,GellMann:1980vs,Mohapatra:1979ia}.
In this model, right-handed neutrinos, whose mass, $M$, is much larger
than the electroweak scale, $M_W$, are coupled to lepton doublets like
right-handed leptons in the SM are. The small ratio $M_W/M$ ensures
the existence of very light mass eigenstates that may be identified
with the observed light neutrinos.  Concerning the baryon asymmetry of
the universe, although the SM contains all the requirements necessary
to dynamically generate the asymmetry, it fails to explain an asymmetry as large
as the one observed~\cite{Dolgov:1991fr}, and now accurately
determined by cosmic microwave background anisotropy
measurements~\cite{Komatsu:2008hk}.  Baryogenesis through
leptogenesis in the original formulation of~\cite{Fukugita:1986hr} is a possible mechanism to explain
the baryon asymmetry.  In this scenario, heavy right-handed neutrinos provide 
both a source of lepton number and CP violation, moreover, they can be out of
equilibrium at temperatures where the SM particles are still thermalized.
Finally, together with many other candidates~\cite{Bertone:2004pz},
light right-handed neutrinos, minimally coupled to SM
particles like in the seesaw mechanism, may provide suitable
candidates for dark-matter particles~\cite{Boyarsky:2009ix}.

Heavy right-handed neutrino play therefore a crucial role in models
trying to explain the neutrino masses and mass hierarchy, and in
leptogenesis.  They may also constitute the heavy partners of light
neutrino families responsible for dark matter. What qualifies a
neutrino as heavy in this context is that its mass is much larger than
the electroweak scale, and consequently of any SM particle. 
This allows for a temperature window in the early universe, where
the temperature is larger than the electroweak scale, but much smaller
than the neutrino mass.  In this temperature window the heavy neutrino
is out of equilibrium, and therefore contributing to the lepton
asymmetry of the universe, while the SM particles may be seen as part
of an in-equilibrium plasma at a temperature $T$.  For such 
temperatures the relevant hierarchy of energy scales is
\begin{equation}
M \gg T \, \gg M_W\,.
\label{scales}
\end{equation}   

The hierarchy of energy scales \eqref{scales} calls for a non-relativistic 
treatment of the heavy neutrino. Because right-handed neutrinos can be embedded 
into Majorana neutrinos, we may want to construct a non-relativistic effective field theory (EFT)
for Majorana fermions along the same line as a non-relativistic EFT for heavy quarks, 
the heavy quark effective theory (HQET), has been built for Dirac fermions \cite{Isgur:1989vq,Eichten:1989zv}.
The construction of this EFT will be the subject of the first part of the paper.
An analogous study can be found in~\cite{Kopp:2011gg}.
The advantages of an EFT treatment for heavy particles over exploiting 
the hierarchy \eqref{scales} in the course of fully relativistic calculations 
in thermal field theory are manifold. First, the EFT makes manifest, already at the 
Lagrangian level, the non-relativistic nature of the Majorana particle.
Second, it allows to separate the computation of relativistic and thermal corrections: 
relativistic corrections are computed setting $T=0$ and contribute to 
the Wilson coefficients of the EFT, whereas thermal corrections are computed in the EFT as small 
corrections affecting the propagation of the non-relativistic Majorana particles in the plasma.
Finally, as we will see, the power counting of the EFT allows a rather 
transparent organization of the calculation leading to several simplifications 
that would not be obvious at the level of the relativistic thermal field theory.

As an application and non-trivial test of the effective field theory, in the second part of the paper 
we compute the thermal corrections to the decay rate of a non-relativistic Majorana neutrino 
within a hot plasma at first order in the SM couplings and at order $T^4/M^3$.
This calculation has been recently done in~\cite{Salvio:2011sf,Laine:2011pq}, 
but with different methods. We will reproduce their result.
In both cases, relativistic thermal field theories have been employed: in~\cite{Salvio:2011sf} 
in the so-called real-time formalism, while in~\cite{Laine:2011pq} in the imaginary-time formalism. 
Here we will use the real-time formalism.\footnote{
Eventually, at the accuracy of this work, the choice between the two formalisms will only  
affect the way thermal condensates are calculated.
}  
It comes as a great simplification within the non-relativistic EFT  
that we will not have to deal with the doubling of degrees of freedom typical of a  
real-time relativistic field theory. As we will see, the thermal calculation becomes 
trivial, while all the computational effort goes into the one loop matching of the EFT, 
which may be performed at zero temperature. 
In the future, the EFT presented here can be used to simplify computations of the decay
rates taking into account CP violation and a medium out of thermal equilibrium, 
as well as for studies of thermal effects in other models 
in which non-relativistic Majorana particles play a role.

The paper is organized as follows. In section~\ref{sec_maj}, we discuss the
non-relativistic degrees of freedom of the EFT. In section~\ref{sec_lep}, we review
some relevant aspects of the model that we use to calculate the Majorana neutrino 
thermal width. In section~\ref{sec_EFT} we derive the relevant EFT Lagrangian. 
Its one-loop Wilson coefficients are calculated in appendix~\ref{appA}. 
The thermal corrections to the width and the final result are presented 
in section~\ref{sec_the}. Some conclusions can be found in section~\ref{sec_con}.

\section{Non-relativistic Majorana fermions}
\label{sec_maj}
In this section, we derive some general properties of a free Majorana fermion in 
the limit where its mass $M$ is much larger than the energy and momentum of any other 
particle in the system. Our aim is to identify the low-energy modes,
write the Majorana free propagator and construct the corresponding Lagrangian.
Low-energy modes are those that may be excited at energies below $M$. 
In the next sections, we will identify the Majorana fermion studied here with a Majorana neutrino, 
and the low-energy degrees of freedom with the low-energy modes 
of the neutrino and the SM particles.

If $\psi$ is a spinor describing a relativistic Majorana particle, then 
\begin{equation}
\psi=\psi^{c}=C  \bar{\psi}^{\,T} \, ,
\label{eq:Majodef}
\end{equation}
where $\psi^{c}$ denotes the charge-conjugate spinor and $C$  
the charge-conjugation matrix that satisfies $C^\dagger=C^T=C^{-1}=-C$ and $C\,\gamma^{\mu\,T}\,C = \gamma^\mu$.\footnote{
A possible choice for $C$ is $C=-i\gamma^{2}\gamma^{0}$.} 
Thus a Majorana spinor has only two independent components.
It is different from a Dirac spinor that has instead four independent components
corresponding to a distinguishable particle and antiparticle.
The relativistic propagators for a free Majorana particle are:
\begin{eqnarray}
\langle 0 | T( \psi^{\alpha} (x) \bar{\psi}^{\beta} (y)  )| 0 \rangle &=& 
i \int \frac{d^{4}p}{(2 \pi)^{4}} \, \frac{(\slashed{p}+M)^{\alpha \beta}}{p^{2}-M^{2}+i\epsilon}  \, e^{-ip \cdot (x-y)} \,,
\label{eq5}
\\
\langle 0 | T( \psi^{\alpha}(x) \psi^{\beta} (y) )| 0 \rangle &=& 
-i \int \frac{d^{4}p}{(2 \pi)^{4}} \, \frac{\left[  (\slashed{p}+M) C \right]^{\alpha \beta}  }{p^{2}-M^{2}+i\epsilon} \,  e^{-ip \cdot (x-y)} \,,
\label{eq6}
\\
\langle 0 | T( \bar{\psi}^{\alpha} (x) \bar{\psi}^{\beta} (y)  ) | 0 \rangle &=& 
-i \int \frac{d^{4}p}{(2 \pi)^{4}} \, \frac{ \left[  C (\slashed{p}+M) \right]^{\alpha \beta}}{p^{2}-M^{2}+i\epsilon} \, e^{-ip \cdot (x-y)} \,,
\label{eq7}
\end{eqnarray}
where $\alpha$ and $\beta$ are Lorentz indices and $T$ stands for the time-ordered product.
Note that, due to the Majorana nature of the fermions and at variance with the Dirac fermion case, 
the combinations $\langle 0 | \psi\psi| 0 \rangle $ and $\langle 0 | \bar{\psi}\bar{\psi}| 0 \rangle$ 
do not vanish. This is a feature that has to be accounted for in the 
relativistic theory when computing amplitudes, since Majorana fields 
may be contracted with vertices involving either particle or antiparticle fields.

In order to identify the low-energy modes of a heavy Majorana field, $\psi$,  
let us assume first that $\psi$, rather than a Majorana field, is a Dirac field describing a heavy quark.
Low-energy modes of a non-relativistic Dirac field have been studied in the framework of HQET~\cite{Neubert:1993mb}. 
In a given reference frame, the momentum of a non-relativistic heavy quark of mass $M$ is $Mv^\mu$, where $v^2=1$,  
up to fluctuations whose momenta, $k^\mu$, are much smaller than $M$. These fluctuations may come 
from the interactions with other particles that, by assumption, carry energies and momenta 
much smaller than $M$. The Dirac field describing a heavy quark can be split 
into a large component, $\psi_{>}$, whose energy is of order $M$, 
and a small component, $\psi_{<}$, whose energy is much smaller than $M$:
\begin{equation}
\psi=\left(\frac{1+\slashed{v}}{2} \right) \psi + \left(\frac{1-\slashed{v}}{2} \right) \psi \equiv \psi_{<} + \psi_{>} \,.
\label{psi}
\end{equation}
According to the above definition: $({1+\slashed{v}})/{2} \times \psi_{<} = \psi_{<}$ and $({1-\slashed{v}})/{2} \times \psi_{>} = \psi_{>}$. 
The small component field, $\psi_{<}$, is eventually matched into the field $h$ of HQET.
This is the field, made of two independent components, that describes in HQET 
the low-energy modes of the heavy quark. It satisfies
\begin{equation}
\frac{1+\slashed{v}}{2} h = h  \,.
\end{equation}
The field $h$ annihilates a heavy quark but does not create an antiquark. 
It satisfies the following equal time anti-commutation relations \cite{Dugan:1991ak}:
\begin{eqnarray}
\left\lbrace h^{\alpha}(t, \vec{x}\,) ,h^{\beta}(t, \vec{y}\,) \right\rbrace &=& 
\left\lbrace \bar{h}^{\alpha }(t, \vec{x}\,) , \bar{h}^{\beta }(t, \vec{y}\,) \right\rbrace = 0 \,,
\label{7ab}
\\
\left\lbrace h^{\alpha}(t, \vec{x}\,), \bar{h}^{\beta }(t, \vec{y}\,) \right\rbrace &=&  
\frac{1}{v^0} \left( \frac{1+\slashed{v}}{2} \right)^{\alpha \beta} \delta^{3}(\vec{x}-\vec{y}\,) \,.
\label{7c}
\end{eqnarray}
The charge conjugated of \eqref{psi} is 
\begin{equation}
\psi^c=\left(\frac{1-\slashed{v}}{2} \right)(C\gamma^{0} \psi^{*}_{<}) + \left(\frac{1+\slashed{v}}{2} \right)(C\gamma^{0} \psi^{*}_{>}) \,,
\label{eq9}
\end{equation}
whose small component, $C\gamma^{0} \psi^{*}_{>}$, may be eventually matched into a 
HQET field, made again of two independent components, that describes the low-energy modes of a heavy antiquark. 
Clearly this field is independent from the one describing the heavy quark:
it annihilates a heavy antiquark but does not create a quark.
It satisfies similar equal time anti-commutation relations as the field $h$.

Let us now go back to consider $\psi$ a field describing a heavy Majorana particle whose momentum 
in some reference frame is $Mv^\mu$ up to fluctuations, $k^\mu$, that are much smaller than~$M$.
Like in \eqref{psi} we may decompose the four-component Majorana spinor into a large and a small component. 
From \eqref{eq:Majodef} it follows, however, that in this case \eqref{psi} and \eqref{eq9} describe the same field, hence 
\begin{equation}
\psi_{<}= C\gamma^{0}\psi^{*}_{>} \, , \quad  \psi_{>}= C\gamma^{0}\psi^{*}_{<}\,.
\label{eq9a}
\end{equation}
This implies that the small component of the Majorana particle field 
coincides with the small component of the Majorana antiparticle field.
In the EFT that describes the low-energy modes of non-relativistic Majorana fermions, 
both the particle and antiparticle modes are described by the same field $N$. 
The field $N$ matches $\psi_{<}$ in the fundamental theory and fulfills
\begin{equation}
\frac{1+\slashed{v}}{2} N = N  \,.
\end{equation}
This is consistent with the Majorana nature of the fermion: we cannot
distinguish a particle from its antiparticle. Note that, while in the fundamental theory 
a Majorana fermion and antifermion are described by the same spinor $\psi$ that is self conjugated, 
in the non-relativistic EFT a Majorana fermion and  antifermion are 
described by the same spinor $N$ that is not self conjugated but 
has by construction only two independent components.
Analogously to the field $h$ in HQET, the field $N$ annihilates a heavy Majorana fermion 
(or antifermion). It satisfies the following equal time anti-commutation relations:
\begin{eqnarray}
\left\lbrace N^{\alpha}(t, \vec{x}\,) , N^{\beta}(t, \vec{y}\,) \right\rbrace &=& 
\left\lbrace \bar{N}^{\alpha}(t, \vec{x}\,) , \bar{N}^{\beta}(t, \vec{y}\,) \right\rbrace = 0 \,,
\\
\left\lbrace N^{\alpha}(t, \vec{x}\,) , \bar{N}^{\beta}(t, \vec{y}\,) \right\rbrace &=& 
\frac{1}{v^0} \left( \frac{1+\slashed{v}}{2} \right)^{\alpha \beta} \delta^{3}(\vec{x}-\vec{y}\,) \,,
\end{eqnarray}
which may be also derived from the full relativistic expression of the Majorana 
spinors given in~\cite{Mannheim:1980eb}.
Finally, we provide the expression for the non-relativistic Majorana propagator. 
Starting from eqs. (\ref{eq5})-(\ref{eq7}), projecting on the small components of the Majorana fields 
and putting $p^{\mu}=Mv^{\mu}+k^{\mu}$, where $k^2 \ll M^2$, we obtain in the large $M$ limit 
\begin{equation}
\langle 0 | T ( N^{\alpha}(x) \bar{N}^{\beta}(y) )  | 0 \rangle 
= \left( \frac{1+\slashed{v}}{2}\right)^{\alpha \beta} \int \frac{d^{4}k}{(2 \pi)^{4}} \, e^{-ik(x-y)}\, \frac{i}{v\cdot k +i\epsilon}  \,  ,
\label{effpropagator}
\end{equation} 
whereas the other possible time-ordered combinations vanish as they contain only creation or annihilation operators. 
The corresponding Lagrangian for a free Majorana fermion is like the HQET Lagrangian: 
\begin{equation}
\mathcal{L}^{(0)}_{\hbox{\tiny N}} = \bar{N}\, i v \cdot \partial \, N \,.
\label{effLag00}
\end{equation}
An analysis of heavy Majorana fermions in an EFT framework analogous to the one presented in this section 
can be also found in~\cite{Kopp:2011gg} (see also~\cite{Hill:2011be}).

\section{Thermal leptogenesis and Majorana neutrinos}
\label{sec_lep}
Starting from this section we will assume an extension of the SM that has been 
implemented in several leptogenesis scenarios~\cite{Fukugita:1986hr,Luty:1992un,Buchmuller:2005eh,Davidson:2008bu}.
It consists of the addition to the SM of some sterile neutrinos with at least one of them having 
mass much larger than the electroweak scale.\footnote{
A similar model but with neutrinos not heavier than the electroweak scale is in~\cite{Asaka:2005pn,Asaka:2005an}.
}
If the temperature of the system, $T$, is such that standard thermal 
leptogenesis is efficiently active, then $T$ is also well above the electroweak scale. 
Assuming that we have well separated neutrino masses, the production of a net lepton asymmetry 
starts when the lightest of the sterile neutrinos, whose mass, $M$, is above the electroweak scale,   
decouples from the plasma reaching an out of equilibrium condition. 
This happens when the temperature drops to $T \sim M$. 
During the universe expansion, the sterile neutrino continues to decay in the regime $T<M$. 
For $T<M$ the recombination process is almost absent and a net lepton asymmetry is generated. 
For illustration, we will consider in the following the simplest case of a SM extension involving only one heavy right-handed neutrino. 
The Lagrangian reads~\cite{Asaka:2006rw}:
\begin{equation}
\mathcal{L}=\mathcal{L}_{\hbox{\tiny SM}} 
+ \frac{1}{2} \,\bar{\psi} \,i \slashed{\partial}  \, \psi  - \frac{M}{2} \,\bar{\psi}\psi 
- F_{f}\,\bar{L}_{f} \tilde{\phi} P_{R}\psi  - F^{*}_{f}\,\bar{\psi}P_{L} \tilde{\phi}^{\dagger}  L_{f} \, ,
\label{eq3}
\end{equation} 
where $\psi = \nu_R + \nu_R^c$ is the Majorana field embedding the right-handed neutrino field $\nu_R$, 
$\tilde{\phi}=i \sigma^{2} \, \phi^*$, with $\phi$ the Higgs doublet, and $L_{f}$ are lepton doublets with flavor $f$.
The Majorana neutrino has mass $M$, $F_f$ is a (complex) Yukawa coupling and 
$P_L = (1 - \gamma^5)/2$, $P_R = (1 + \gamma^5)/2$ are the left-handed and right-handed projectors respectively.
Lepton doublets, $L_{f}$, carry SU(2) indices, which are contracted with those of the Higgs doublet, $\phi$,
and Lorentz indices, which are contracted with those carried by the Majorana fields.
Right-handed neutrinos are sterile, hence their interaction has not been gauged.
Because we are considering the Lagrangian \eqref{eq3} for a neutrino mass $M$ and a temperature $T$
much larger than the electroweak scale, the SM Lagrangian, $\mathcal{L}_{\hbox{\tiny SM}}$, 
is symmetric under an unbroken SU(2)$\times$U(1) gauge symmetry and its particles are massless.

\section{Effective field theory for non-relativistic Majorana neutrinos}
\label{sec_EFT}
By construction, an EFT suitable to describe non-relativistic Majorana neutrinos 
must be, under the condition \eqref{scales}, equivalent to our fundamental theory \eqref{eq3} 
order by order in $\Lambda/M$. The scale $\Lambda$ is the ultraviolet cut-off of the EFT and is such that $T \ll \Lambda \ll M$. 
The relevant degrees of freedom at the scale of the temperature are the non-relativistic Majorana field, $N$, 
introduced in section~\ref{sec_maj}, which describes the Majorana neutrino, and the SM particles.
The temperature is well above the electroweak scale.
Hence, the relevant symmetry is an unbroken SU(2)$\times$U(1) gauge symmetry, which 
implies that all particles with the exception of the heavy Majorana neutrino are massless. 
The EFT is written as an expansion in local operators and powers of $1/M$.
The higher the dimension of the operator, the more its contribution to physical observables 
is suppressed by powers of $T/M$. In the following, we will consider only operators 
up to dimension seven, i.e. contributing up to order $1/M^3$ to physical observables. 

The EFT Lagrangian has the general structure 
\begin{equation}
\mathcal{L}_{\hbox{\tiny EFT}}= \mathcal{L}_{\hbox{\tiny SM}}+\mathcal{L}_{\hbox{\tiny N}}+\mathcal{L}_{\hbox{\tiny N-SM}}\,,
\label{eq10}
\end{equation}
where $\mathcal{L}_{\hbox{\tiny N}}$ describes the propagation of the non-relativistic Majorana 
neutrino and $\mathcal{L}_{\hbox{\tiny N-SM}}$ its interaction with the SM particles.
The Lagrangian's parts $\mathcal{L}_{\hbox{\tiny N}}$ and $\mathcal{L}_{\hbox{\tiny N-SM}}$ are determined 
by matching at the scale $\Lambda$ matrix elements in the EFT with matrix elements computed in~\eqref{eq3}.
A crucial observation is that, in the matching, $T$ can be set to zero because $\Lambda \gg T$; 
hence $\mathcal{L}_{\hbox{\tiny EFT}}$ can be computed in the vacuum.
In the following two paragraphs, we will write $\mathcal{L}_{\hbox{\tiny N}}$ and $\mathcal{L}_{\hbox{\tiny N-SM}}$
at the accuracy needed to compute the Majorana neutrino thermal width 
at first order in the SM couplings and at order $T^4/M^3$.
In a given reference frame the momentum of the Majorana neutrino is $Mv^\mu$ up to fluctuations,
which are much smaller than $M$. 

At order $1/M^0$ the Lagrangian $\mathcal{L}_{\hbox{\tiny N}}$ would coincide with \eqref{effLag00}, if the Majorana 
neutrino would be stable at zero temperature. However, the Majorana neutrino may decay into a Higgs and a lepton.
Accounting for this modifies the Lagrangian \eqref{effLag00} into 
\begin{equation}
\mathcal{L}_{\hbox{\tiny N}}= \bar{N} \left(iv \cdot \partial - \frac{i\Gamma^{(0)}_{T=0}}{2} \right)N + \mathcal{O}\left(\frac{1}{M}\right) \,,
\label{effLag0}
\end{equation}
where $\Gamma^{(0)}_{T=0}$ is the decay width at zero temperature in the heavy-mass limit.
It has been computed previously in the literature~\cite{Salvio:2011sf,Laine:2011pq} and reads at leading order 
\begin{equation}
\Gamma^{(0)}_{T=0}=\frac{|F|^2M}{8 \pi} \,, 
\label{dt0}
\end{equation}
where $|F|^2=\sum_{f=1}^3F^*_f F_f$. 

\begin{figure}[htb]
\centering
\includegraphics[scale=0.55]{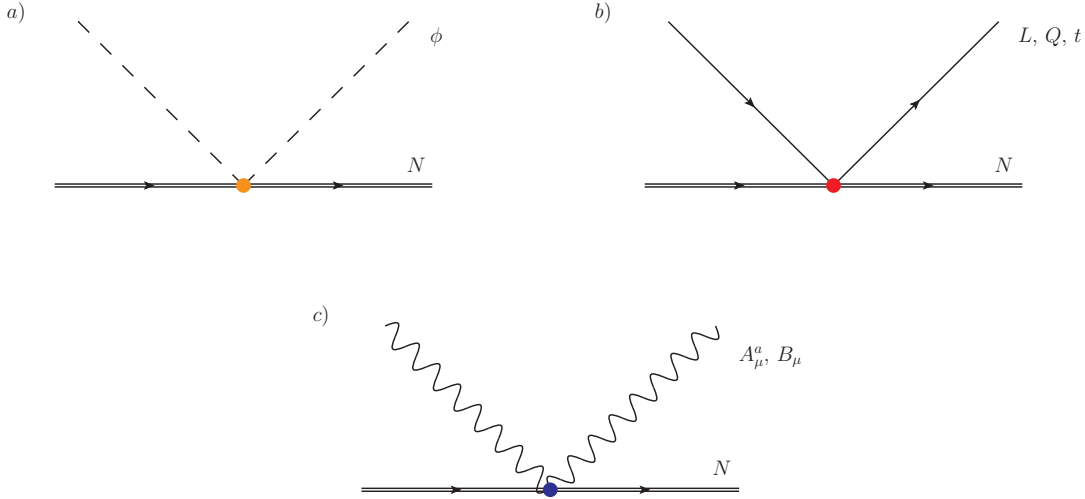}
\caption{\label{Fig1} Diagrams showing the different types of vertices 
induced by the EFT Lagrangian $\mathcal{L}_{\hbox{\tiny N-SM}}$.
These involve interactions between heavy Majorana neutrinos and Higgs fields in $a$, 
fermions in $b$ and the gauge bosons in $c$.}
\end{figure}

The Lagrangian $\mathcal{L}_{\hbox{\tiny N-SM}}$, organized in an expansion in $1/M$, reads
\begin{equation}
\mathcal{L}_{\hbox{\tiny N-SM}}=\frac{1}{M}\mathcal{L}_{\hbox{\tiny N-SM}}^{(1)}
+\frac{1}{M^2}\mathcal{L}_{\hbox{\tiny N-SM}}^{(2)}
+\frac{1}{M^3}\mathcal{L}_{\hbox{\tiny N-SM}}^{(3)}
+\mathcal{O}\left(\frac{1}{M^4}\right),
\end{equation}
where $\mathcal{L}_{\hbox{\tiny N-SM}}^{(n)}$ includes all operators of dimension $4+n$.
They describe the effective interactions between the Majorana neutrino and 
the Higgs field $\phi$, the lepton doublets $L_f$ of all flavours $f$,
the heavy-quark doublets $Q^T=(t,b)$, where $t$ stands for the top field and $b$ for the bottom field, 
the right-handed top field and the SU(2)$\times$U(1) gauge bosons (see diagrams in figure~\ref{Fig1}). 
We consider only Yukawa couplings with the top quark and neglect Yukawa couplings with other quarks and leptons, 
for the ratio of Yukawa couplings is proportional to the ratio of the corresponding fermion masses
when the gauge symmetry is spontaneously broken.
The number of operators contributing to $\mathcal{L}_{\hbox{\tiny N-SM}}$
may be further significantly reduced by assuming the Majorana neutrino at rest and 
by selecting only operators that could contribute to the Majorana neutrino thermal width 
at first order in the SM couplings and at order $T^4/M^3$.
At first order in the SM couplings, thermal corrections are encoded into tadpole diagrams. 
Hence we need to consider only operators with imaginary coefficients (tadpoles do not develop an imaginary part), 
made of two Majorana fields with no derivatives acting on them (the Majorana neutrino is at rest), 
coupled to bosonic operators with an even number of spatial and time derivatives 
(the boson propagator in the tadpole is even for space and time reflections) 
and to fermionic operators with an odd number of derivatives 
(the massless fermion propagator in the tadpole is odd for spacetime reflections).
Finally, we may use field redefinitions to get rid of operators containing terms like 
$\slashed{\partial}$(fermion field) or $\partial^2$(boson field). 
The Lagrangian $\mathcal{L}_{\hbox{\tiny N-SM}}^{(1)}$ reads 
\begin{equation}
\mathcal{L}_{\hbox{\tiny N-SM}}^{(1)}=a \; \bar{N} N \, \phi^{\dagger} \phi \, .
\label{Wa}
\end{equation}
The Lagrangian $\mathcal{L}_{\hbox{\tiny N-SM}}^{(2)}$ does not contribute to our observable 
because it involves either boson fields with one derivative or fermion fields with no derivatives.
The Lagrangian $\mathcal{L}_{\hbox{\tiny N-SM}}^{(3)}$ reads
\begin{eqnarray}
\mathcal{L}_{\hbox{\tiny N-SM}}^{(3)} &=&  
~~ b \; \bar{N}N \,  \big(v\cdot D \phi^{\dagger}\big)  \,  \big(v\cdot D \phi\big) 
\nonumber  \\ 
&& + c^{ff'}_{1} \; \left[ \left(\bar{N}P_{L} \, iv\cdot D L_{f}\right) \left(\bar{L}_{f'} P_{R} N \right) \right.
\nonumber \\
&&
\left.
\hspace{1cm}  
+ \left( \bar{N} P_{R} \, i v\cdot D L^{c}_{f'} \right) \left( \bar{L}^{c}_{f} P_{L}  N \right)  \right]  
\nonumber \\
&& 
+ c^{ff'}_{2} \; \left[  \left(\bar{N} P_{L} \, \gamma_{\mu} \gamma_{\nu} \, iv\cdot D L_{f} \right) 
\left( \bar{L}_{f'} \, \gamma^{\nu} \gamma^{\mu} \, P_{R} N \right)  \right.
\nonumber \\
&&
\left.
\hspace{1cm}  
+ \left(\bar{N}  P_{R} \, \gamma_{\mu} \gamma_{\nu} \, iv\cdot D L^{c}_{f'} \right) 
\left( \bar{L}^{c}_{f} \, \gamma^{\nu} \gamma^{\mu} \, P_{L} N \right) \right]  
\nonumber \\ 
&& 
+ c_{3} \; \bar{N}N \, \left( \bar{t}P_{L} \, v^\mu v^\nu \gamma_{\mu} \, i D_{\nu} t \right) 
+ c_{4} \; \bar{N}N \, \left( \bar{Q}P_{R} \, v^\mu v^\nu \gamma_{\mu} \, i D_{\nu} Q \right)
\nonumber \\
&& 
+ c_{5} \; \bar{N}\,\gamma^5\gamma^\mu\,N \, \left( \bar{t}P_{L} \, v\cdot\gamma \, i D_{\mu} t \right) 
+ c_{6} \; \bar{N}\,\gamma^5\gamma^\mu\,N \, \left( \bar{Q}P_{R} \, v\cdot\gamma \, i D_{\mu} Q \right) 
\nonumber \\
&& 
+ c_{7} \; \bar{N}\,\gamma^5\gamma^\mu\,N \, \left( \bar{t}P_{L} \, \gamma_\mu \, i v\cdot D t \right)
+ c_{8} \; \bar{N}\,\gamma^5\gamma^\mu\,N \, \left( \bar{Q}P_{R} \, \gamma_\mu \, i v\cdot D Q \right)
\nonumber \\
&& 
- d_{1} \; \bar{N}N \,  v^\mu v_\nu W^{a}_{\alpha\mu} W^{a\,\alpha\nu}  \
- d_{2} \; \bar{N}N \,  v^\mu v_\nu F_{\alpha\mu} F^{\alpha\nu} 
\nonumber \\
&&
+ d_{3} \; \bar{N}N \, W^{a}_{\mu\nu} W^{a\,\mu\nu}  \
+ d_{4} \; \bar{N}N \, F_{\mu\nu} F^{\mu\nu} 
\, .
\label{Wb}
\end{eqnarray}
The fields $W^a_{\mu\nu}$ and $F_{\mu\nu}$ are the field strength tensors of the SU(2) 
gauge fields, $A^a_{\mu}$, and U(1) gauge fields, $B_{\mu}$, respectively.
For the operators multiplying $c^{ff'}_{1}$ and $c^{ff'}_{2}$
the SU(2) indices of $L_{f}$ and $\bar{L}_{f'}$ are contracted with each other 
while their Lorentz indices are contracted with gamma matrices and Majorana fields.

The Wilson coefficients $a$, $b$, $c^{ff'}_{i}$, $c_i$ and $d_i$ encode all contributions coming from the high-energy modes 
of order $M$ that have been integrated out when matching from the fundamental theory \eqref{eq3} to the
EFT  \eqref{eq10}. We are interested only in their imaginary parts. At first order in the SM couplings they read
\begin{eqnarray}
&& {\rm{Im}}\,a = -\frac{3}{8\pi}|F|^{2}\lambda \,, 
\label{coa}\\
&& {\rm{Im}}\,b =- \frac{5}{32 \pi} (3g^{2}+g'^{\,2})|F|^{2}\,, 
\label{cob}\\
&& {\rm{Im}}\,c^{ff'}_{1} = \frac{3}{8\pi}|\lambda_{t}|^{2}F_{f'} F^{*}_{f} -  \frac{3}{16 \pi}(3g^{2}+g'^{\,2})F_{f'} F^{*}_{f}  \,, 
\label{co1}\\
&& {\rm{Im}}\,c^{ff'}_{2} = \frac{1}{384 \pi} (3g^{2}+g'^{\,2})F_{f'} F^{*}_{f}  \,,
\label{co2}\\
&& {\rm{Im}}\,c_{3} = \frac{1}{24\pi}|\lambda_{t}|^{2}|F|^{2} \,, 
\qquad {\rm{Im}}\,c_{4} = \frac{1}{48\pi}|\lambda_{t}|^{2}|F|^{2} \,,
\label{co34}\\
&& {\rm{Im}}\,c_{5} = \frac{1}{48\pi}|\lambda_{t}|^{2}|F|^{2} \,,
\qquad {\rm{Im}}\,c_{6} = \frac{1}{96\pi}|\lambda_{t}|^{2}|F|^{2} \,,
\label{co56}\\
&& {\rm{Im}}\,c_{7} = \frac{1}{48\pi}|\lambda_{t}|^{2}|F|^{2} \,,
\qquad {\rm{Im}}\,c_{8} = \frac{1}{96\pi}|\lambda_{t}|^{2}|F|^{2} \,,
\label{co78}\\
&& {\rm{Im}}\,d_{1} = - \frac{1}{96\pi}g^{2}|F|^{2} \,, 
\qquad {\rm{Im}}\,d_{2} =- \frac{1}{96\pi}g'^{\,2}|F|^{2}\,,
\label{cod12}\\
&& {\rm{Im}}\,d_{3} = - \frac{1}{384\pi}g^{2}|F|^{2} \,, 
\qquad\!\!\! {\rm{Im}}\,d_{4} =- \frac{1}{384\pi}g'^{\,2}|F|^{2}\,,
\label{cod34}
\end{eqnarray}
where $g$ is the SU(2) coupling, $g'$ the U(1) coupling, 
$\lambda$ the four-Higgs coupling and $\lambda_t$ the top Yukawa coupling.
We refer to appendix~\ref{appA} for details on the calculation.

If the Majorana neutrino is not at rest, then we need to add to \eqref{Wb} 
operators that depend on the neutrino momentum. The leading operator is the dimension seven operator 
\begin{equation}
- \frac{1}{2 M^3} a \; \bar{N} \left[\partial^2-(v\cdot \partial)^2\right] N \, \phi^{\dagger} \phi \, .
\label{Wmomdep}
\end{equation}
The Wilson coefficient of this operator is fixed by the relativistic dispersion relation
\begin{equation}
\bar{N}N\, \left(\sqrt{(M+\delta m)^2+\vec{k}^{\,2}} - M \right) = 
\bar{N}N\,\left(\delta m + \frac{\vec{k}^{\,2}}{2M} - \delta m\, \frac{\vec{k}^{\,2}}{2M^2} + \dots\right) \,,
\end{equation}
with $\delta m = - a\,\phi^{\dagger} \phi/M$, or by methods similar to those developed in~\cite{Brambilla:2003nt}.

The EFT Lagrangian derived in this section follows from symmetry arguments and standard (one-loop) perturbation theory.  
Owing to the hierarchy \eqref{scales}, the temperature could be set to zero when computing the Wilson coefficients.
Thermal effects factorize. We consider this factorization the main advantage in the use of the EFT. 
Moreover, the calculation of the Majorana neutrino thermal width will turn out to be very simple. 
Indeed, already at this level, the structure and power counting of the EFT allow to make some general statements 
about the origin and size of the different contributions. The width will be the sum of contributions coming 
from the scattering with Higgs, SM fermions (either leptons or left-handed heavy quarks or right-handed tops) 
and gauge fields in the early universe plasma. We call these contributions 
$\Gamma_{\phi}$, $\Gamma_{\rm fermions}$ and $\Gamma_{\rm gauge}$ respectively.
The leading operator responsible for the interaction of the Majorana neutrino with the Higgs 
is  the dimension five operator \eqref{Wa}, hence the natural power counting of the EFT implies 
\begin{equation}
\Gamma_{\phi} \sim \frac{T^{2}}{M}  \,.
\label{eq0}
\end{equation}
This is also the leading contribution to the thermal width of the Majorana neutrino.
The interaction of the Majorana neutrino with the SM fermions and the gauge bosons is mediated 
in \eqref{Wb} by operators of dimension seven, hence 
\begin{equation}
\Gamma_{\hbox{\tiny fermions}} \sim \frac{T^{4}}{M^{3}} \,, \qquad 
\Gamma_{\hbox{\tiny gauge}} \sim \frac{T^{4}}{M^{3}} \, .
\label{eq}
\end{equation}
In the next section, we will compute $\Gamma_{\phi}$, $\Gamma_{\hbox{\tiny fermions}}$ and $\Gamma_{\hbox{\tiny gauge}}$
at first order in the SM couplings.

\section{Thermal width}
\label{sec_the}
\begin{figure}[htb]
\centering
\includegraphics[scale=0.6]{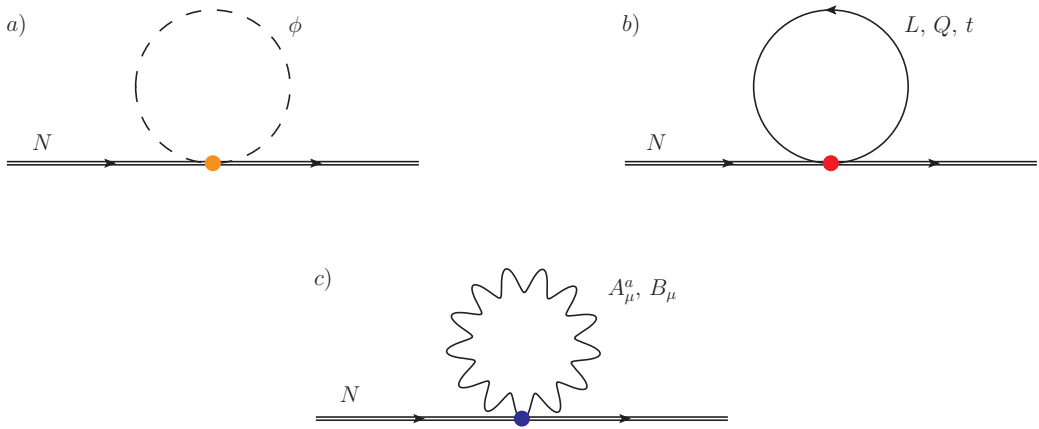}
\caption{\label{Fig2} Tadpole diagrams contributing to the thermal
width of a heavy Majorana neutrino at first order in the SM couplings.
The heavy Majorana neutrino is represented by a double line, the
Higgs propagator by a dashed line, fermion propagators (leptons, heavy
quark doublets and top singlet) by a continuous line and gauge bosons by
a wiggled line.}
\end{figure}

A Majorana neutrino in a plasma of SM particles thermalized at some temperature $T$ decays with a width 
$\Gamma = \Gamma_{T=0}+\Gamma_T$, where $\Gamma_{T=0}$ is the in-vacuum width and 
$\Gamma_{T}$ encodes the thermal corrections to the width induced by the interaction with the particles in the plasma.
We call $\Gamma_{T}$ the Majorana neutrino thermal width.
The decay of the Majorana neutrino happens at a distance of order $1/M$. The neutrino 
releases a large amount of energy of the order of its mass into a   
Higgs and lepton pair. The interaction vertex is described by the Lagrangian \eqref{eq3}.
At such small distances the neutrino is insensitive to the plasma and the decay 
happens as in the vacuum. The width is $\Gamma_{T=0}$, which at leading order can be read off eq. \eqref{dt0}.\footnote{
Next-to-leading order corrections in the SM couplings to $\Gamma_{T=0}$ have been calculated in~\cite{Salvio:2011sf,Laine:2011pq}.
Those corrections may be taken over in the EFT to improve the expression 
of the zero-temperature Majorana neutrino width in $\mathcal{L}_{\hbox{\tiny N}}$.
}
At distances of order $1/T$, the vertices involving Majorana neutrinos in the fundamental Lagrangian \eqref{eq3}
cannot be resolved, instead the Majorana neutrino effectively interacts with Higgs, 
fermion and gauge boson pairs as shown in figure~\ref{Fig1}. 
These are the vertices in the EFT that can be read off eqs.~\eqref{Wa} and \eqref{Wb}.
The effective couplings of these vertices are the Wilson coefficients listed in \eqref{coa}-\eqref{cod34}.
They are all of first order in the SM couplings $g^2$, $g'^{\,2}$, $\lambda$ and $|\lambda_t|^2$. 
Hence, at that order, only tadpole diagrams of the type shown in 
figure~\ref{Fig2} can contribute to the Majorana neutrino width.
Tadpoles do not vanish (in dimensional regularization) only if the momentum circulating in the loop is of the order 
of the plasma temperature, instead  they induce a thermal correction, $\Gamma_T$, to the width.
In the following, we will calculate $\Gamma_T$ assuming that the thermal bath 
of SM particles is at rest with respect to the Majorana neutrino.
Moreover, we choose our reference frame such that $v^\mu = (1,\vec{0}\,)$. 

We calculate finite temperature effects in the so-called real-time formalism. 
This amounts at modifying the contour of the time integration in the partition function to allow for real time.
The modified contour has two lines stretching along the real-time axis. 
A consequence of this is that in the real-time formalism the degrees of freedom double. 
One usually refers to them as degrees of freedom 
of type $1$ and $2$. The physical degrees of freedom, those describing 
initial and final states, are of type $1$. Propagators can mix fields 
of type $1$ with fields of type $2$, while vertices do not couple 
fields of different types. We refer to textbooks like~\cite{LeBellac:1996} for more details.
It has been shown in~\cite{Brambilla:2008cx} that because the $12$ component of a 
heavy-field propagator vanishes in the heavy-mass limit, heavy fields 
of type $2$ decouple from the theory and can be neglected.  
This also applies to the Majorana neutrino field $N$, 
which may be considered of type $1$ only.  
In our case, we will calculate the tadpole diagrams shown in figure~\ref{Fig2}. 
Because there the SM fields couple directly to the neutrino field $N$, 
also the SM fields may be considered to be of type $1$ only.
This is a significant simplification in the calculation that the non-relativistic 
EFT makes manifest from the beginning.

Tadpole diagrams like those shown in figure~\ref{Fig2} involve only $11$ components 
of the real-time propagators of the SM fields. 
The $11$ component is the time-ordered propagator of the physical field; 
for a bosonic (scalar) field propagating from~$0$ to~$x$ it reads  
\begin{equation}
i\Delta(x)= \int \frac{d^{4}q}{(2 \pi)^{4}} \,e^{-i q \cdot x}\left[ \frac{i}{q^{2}+i\epsilon } 
+ 2\pi n_{\hbox{\tiny B}}(|q_{0}|)\delta(q^{2}) \right] \,,
\label{eq12}
\end{equation}
where  $n_{\hbox{\tiny B}}(|q_{0}|)=1/(e^{|q_{0}|/T}-1)$ is the Bose--Einstein distribution in the rest frame, 
and for a fermionic field propagating from $0$ to $x$ 
\begin{equation}
iS(x)= \int \frac{d^{4}q}{(2 \pi)^{4}}\,e^{-i q \cdot x}\,\slashed{q} \left[ \frac{i}{q^{2}+i\epsilon } 
- 2\pi n_{\hbox{\tiny F}}(|q_{0}|)\delta(q^{2}) \right] \,,
\label{eq13}
\end{equation}
where  $n_{\hbox{\tiny F}}(|q_{0}|)=1/(e^{|q_{0}|/T}+1)$ is the Fermi--Dirac distribution in the rest frame.
The first terms in \eqref{eq12} and \eqref{eq13}  are the in-vacuum propagators. 
We recall that SM particles are massless in the high-temperature regime \eqref{scales}.

Thermal corrections to the decay width can be computed from the Majorana neutrino propagator in momentum space:
\begin{equation}
\int d^{4}x\,\,e^{i k \cdot x}\, \langle T(N^{\alpha}(x) N^{\dagger\,\beta }(0)) \rangle_T^{\hbox{\tiny int}} \,,
\label{fullNN}
\end{equation}
where $\langle \cdots \rangle_T^{\rm int}$ stands for the thermal average evaluated on the action
$\displaystyle \int d^4x\,\mathcal{L}_{\hbox{\tiny EFT}}$.
In the $v^\mu = (1,\vec{0}\,)$ frame, the Majorana neutrino propagator has the general form (cf. with \eqref{effpropagator})
\begin{equation}
\left( \frac{1+\gamma_0}{2}\right)^{\alpha\beta} \frac{iZ}{k^{0}- E+i{\Gamma}/{2}}=
\left( \frac{1+\gamma_0}{2}\right)^{\alpha\beta} Z\left[ 
\frac{i}{k^{0}+i\epsilon} - \left(i E+\frac{\Gamma}{2}\right)\left( \frac{i}{k^{0}+i\epsilon} \right)^{2}  
+  \cdots \right].
\label{eq15}
\end{equation}
The wavefunction normalization $Z$, mass shift $E$ and width $\Gamma$ are determined by self-energy diagrams.
In our case, we consider only the tadpole diagrams shown in figure~\ref{Fig2}. 
Because $Z-1$ is given by the derivative of the self-energy with respect to the incoming momentum 
and because tadpole diagrams do not depend on the incoming momentum, we have that $Z=1$.
In the expansion \eqref{eq15}, the width $\Gamma$ is then twice the real part of the residue of the double pole in $k^0=0$. 

We start by considering the contribution to the decay width from the Higgs tadpole 
(diagram $a$ in figure~\ref{Fig2}). A Higgs tadpole may contribute to \eqref{fullNN} either 
through the dimension five operator \eqref{Wa} or through the dimension seven operator in the 
first line of \eqref{Wb} or through higher-order operators. 
Expanding \eqref{fullNN} in $\mathcal{L}_{\hbox{\tiny N-SM}}$, we obtain
\begin{eqnarray}
&& i \frac{a}{M} \int d^{4}x\,e^{i k \cdot x}\, \langle \int d^{4}z\, T( N^{\alpha}(x) N^{\dagger \,\beta}(0) \, 
N^{\dagger \,\mu}(z) N_{\mu}(z) \phi^{\dagger}(z) \phi(z) )\rangle_T^{\hbox{\tiny free}} 
\nonumber\\
&+& i \frac{b}{M^3} \int d^{4}x\,\,e^{i k \cdot x}\, \langle \int d^{4}z\, T( N^{\alpha}(x) N^{\dagger \,\beta}(0) \, 
N^{\dagger \,\mu}(z) N_{\mu}(z) \partial_0\phi^{\dagger}(z) \partial_0\phi(z) )\rangle_T^{\hbox{\tiny free}} 
\nonumber\\
&+& \hbox{contributions of higher order in}~1/M\,, 
\label{gammaab}
\end{eqnarray}
where $\langle \cdots \rangle_T^{\hbox{\tiny free}}$ stands for the thermal average evaluated on the action
$\displaystyle \int d^4x\,(\mathcal{L}_{\hbox{\tiny SM}}+\mathcal{L}_{\hbox{\tiny N}})$.
The Wilson coefficients $a$ and $b$ can be read off eqs. \eqref{coa} and \eqref{cob} respectively.
Because the Majorana neutrinos do not thermalize, we have that 
\begin{equation}
\langle \hbox{(Majorana fields)} \times \hbox{(SM fields)}\rangle_T^{\hbox{\tiny free}} = 
\langle 0|\hbox{(Majorana fields)} |0\rangle \times \langle \hbox{(SM fields)}\rangle_T\,,
\end{equation}
where $\langle 0|\hbox{(Majorana fields)} |0\rangle$ is a free Green's function that can be  
computed by contracting the Majorana neutrino fields according to \eqref{effpropagator}, and 
$\langle \cdots \rangle_T$ is a thermal average of SM fields weighted by the SM partition function.
Comparing \eqref{gammaab} with \eqref{eq15}, we obtain 
\begin{eqnarray}
\Gamma_{\phi} &=& 2 \frac{{\rm Im} \, a}{M} \langle \phi^{\dagger}(0) \phi(0) \rangle_{T}  
+ 2 \frac{{\rm Im} \, b}{M^3} \langle \partial_0\phi^{\dagger}(0) \partial_0\phi(0) \rangle_{T}  
\nonumber\\
&=&  \frac{{\rm Im} \, a}{3} \frac{T^2}{M} +  \frac{2\pi^2}{15} {\rm Im} \, b \frac{T^4}{M^3}\,.
\label{gammaphi}
\end{eqnarray}
The last line follows from having computed the Higgs thermal condensates at leading order:
\begin{eqnarray}
\langle \phi^{\dagger}(0) \phi(0) \rangle_{T}  &=&  
2 \int  \frac{d^{4}q}{(2\pi)^{4}}\, 2\pi n_{\hbox{\tiny B}}(|q_{0}|)\delta(q^{2})  = \frac{T^2}{6}\,,
\label{cond1}\\
\langle  \partial_0\phi^{\dagger}(0) \partial_0\phi(0) \rangle_{T}  &=&  
2 \int  \frac{d^{4} q}{(2\pi)^{4}} \, q_0^2 \, 2\pi n_{\hbox{\tiny B}}(|q_{0}|)\delta(q^{2})  = \frac{\pi^2}{15}T^4\,.
\label{cond2}
\end{eqnarray}
We have used dimensional regularization to get rid of the vacuum contributions. 
We observe that bosonic condensates involving an odd number of spatial or time derivatives  
would give rise to vanishing momentum integrals. 

If the Majorana neutrino is not at rest, the operator \eqref{Wmomdep} induces a momentum 
dependent correction. It reads
\begin{eqnarray}
\Gamma_{\phi,\hbox{\tiny mom.\,dep.}} &=& 
2 \frac{{\rm Im} \, a}{M} \left(-\frac{\vec{k}^{\,2}}{2M^2} \right) \langle \phi^{\dagger}(0) \phi(0) \rangle_{T}  
= -  \frac{{\rm Im} \, a}{6}\frac{\vec{k}^{\,2}T^2}{M^3} \,.
\label{gammamomdep}
\end{eqnarray}

In a similar way we can compute the contribution to the decay width from the fermion tadpoles
(diagram $b$ in figure~\ref{Fig2}):
\begin{eqnarray}
\Gamma_{\hbox{\tiny fermions}} &=& - \left(\frac{{\rm Im}\, c^{ff'}_{1}}{2M^3} + \frac{ 2 {\rm Im}\, c^{ff'}_{2} }{M^3} \right)
\langle \bar{L}_{f'}(0)\gamma^0 iD_0 L_f(0) \rangle_{T}  
\nonumber\\
&& + 2 \frac{{\rm Im}\, c_{3}}{M^3} \langle \bar{t}(0) P_L\gamma^0 iD_0 t(0) \rangle_{T}  
+ 2 \frac{{\rm Im}\, c_{4}}{M^3} \langle \bar{Q}(0) P_R\gamma^0 iD_0 Q(0) \rangle_{T}  
\nonumber\\
&=&  \left( -{\rm Im}\, c^{ff}_{1} -4 {\rm Im}\, c^{ff}_{2}  + 3 {\rm Im}\, c_{3} + 6 {\rm Im}\, c_{4} \right) 
\frac{7 \pi^2}{60} \frac{T^4}{M^3}\,,
\label{gammafermions}
\end{eqnarray}
where the Wilson coefficients $c^{ff}_{i}$ and $c_i$ can be read off eqs. \eqref{co1}-\eqref{co34}.
The last line of \eqref{gammafermions} follows from having computed the lepton thermal condensate at leading order, 
\begin{equation}
\langle \bar{L}_{f'}(0)\gamma^0 iD_0 L_f(0) \rangle_{T} =
-2 \delta_{ff'} \int  \frac{d^{4}q}{(2\pi)^{4}}\, q_0 \, {\rm Tr}\, \left\{ \gamma^0\slashed{q} \right\} 
\, (-2\pi) n_{\hbox{\tiny F}}(|q_{0}|)\delta(q^{2})  = \frac{7 \pi^2}{30} T^4\,,
\label{cond3}
\end{equation}
and similarly the quark condensates, $\langle \bar{t}(0) P_L\gamma^0 iD_0 t(0) \rangle_{T}  = 7 \pi^2 T^4/40$ 
and $\langle \bar{Q}(0) P_R\gamma^0$ $\times iD_0 Q(0) \rangle_{T} = 7 \pi^2 T^4/20$.
We note that fermionic condensates involving an even number of derivatives would give rise to vanishing momentum integrals. 

Tadpole diagrams generated by operators multiplying the Wilson coefficients $c_5$, $c_6$, $c_7$ and $c_8$ 
in \eqref{Wb} provide a contribution to the width that depends on the spin coupling of the Majorana neutrino with the medium.\footnote{
The operator $N^{\dagger}\,\gamma^5\gamma^i\,N$ can be also written as $-2 N^{\dagger}\,S^i\,N$, where $\vec{S}$ is the spin 
operator.} If the medium is isotropic, this coupling is zero.

Finally, the contribution to the decay width from the gauge boson tadpoles (diagram $c$ in figure~\ref{Fig2}) gives
\begin{eqnarray}
\Gamma_{\hbox{\tiny gauge}} &=&  2 \frac{{\rm Im}\, d_{1}}{M^3} \langle W^a_{0i}(0) W^a_{0i}(0) \rangle_{T}  
+ 2 \frac{{\rm Im}\, d_{2}}{M^3} \langle F_{0i}(0) F_{0i}(0) \rangle_{T}  
\nonumber\\
&=&  \left( 3 {\rm Im}\, d_{1} +  {\rm Im}\, d_{2} \right) \frac{2 \pi^2}{15} \frac{T^4}{M^3}\,,
\label{gammagauge}
\end{eqnarray}
where the Wilson coefficients $d_{i}$ can be read off eq. \eqref{cod12}.
The last line of \eqref{gammagauge} follows from having computed the gauge boson thermal electric condensates 
at leading order~\cite{Brambilla:2008cx}: 
$\langle W^a_{0i}(0) W^a_{0i}(0) \rangle_{T}   = \pi^2 T^4/5$ 
and $ \langle F_{0i}(0) F_{0i}(0) \rangle_{T} =  \pi^2 T^4/15$.
The operators $\bar{N}N \, W^{a}_{\mu\nu} W^{a\,\mu\nu}$ and $\bar{N}N \, F_{\mu\nu} F^{\mu\nu}$ in the last line
of \eqref{Wb} do not contribute to the thermal width because at leading order    
$\langle W^a_{\mu\nu}(0) W^{a\,\mu\nu}(0) \rangle_{T}   = \langle F_{\mu\nu}(0) F^{\mu\nu}(0) \rangle_{T} = 0$.

The above expressions for the thermal decay widths induced by Higgs, fermions and gauge bosons 
are consistent with the estimates \eqref{eq0} and \eqref{eq} obtained by sole power-counting arguments.
Summing up $\Gamma_{\phi}$, $\Gamma_{\phi,\hbox{\tiny mom.\,dep.}}$, $\Gamma_{\hbox{\tiny fermions}}$ and $\Gamma_{\hbox{\tiny gauge}}$ 
and using the explicit expressions of the Wilson coefficients, we get 
at first order in the SM couplings and at order $T^4/M^3$ the Majorana neutrino thermal width:
\begin{equation}
\Gamma_T = 
\frac{|F|^{2}M}{8\pi}\left[-\lambda \left( \frac{T}{M} \right)^{2} 
+ \frac{\lambda}{2} \frac{\vec{k}^{\,2}\,T^2}{M^4}
- \frac{\pi^{2}}{80}\left( \frac{T}{M} \right)^{4}(3g^{2}+g'^{\,2}) 
- \frac{7\pi^{2}}{60}\left( \frac{T}{M} \right)^{4}|\lambda_{t} |^{2} \right] .
\label{eq20}
\end{equation} 
If the neutrino is at rest, we can set $\vec{k}=\vec{0}$.
Equation \eqref{eq20} agrees with the analogous expression derived in~\cite{Salvio:2011sf} 
up to order $T^2/M$. It also agrees with the result of~\cite{Laine:2011pq} up to order $T^4/M^3$.
In~\cite{Laine:2011pq} also corrections of order $\vec{k}^{\,2}T^4/M^5$ have been computed.
We note that we could express our results \eqref{gammaphi}, \eqref{gammamomdep}, \eqref{gammafermions} 
and \eqref{gammagauge} also in terms of Higgs, lepton, quark and gauge field condensates.
This appears to be a straightforward consequence of the EFT, which requires, 
at the order considered here, that thermal corrections are encoded into tadpole diagrams. 
In relation to $\Gamma_T$, condensates have been also discussed in~\cite{Laine:2011pq}.

\section{Conclusions}
\label{sec_con}
In this work we have built an effective field theory 
for non-relativistic Majorana fermions and we have shown 
some advantages of such an approach for computations in a thermal medium. 
The EFT is similar to HQET but keeps track 
of the Majorana nature of the fermion by describing both the 
particle and the antiparticle with the same field.

Although the approach is quite general, as a proof of concept 
we apply it to compute corrections to the Majorana neutrino decay rate 
induced by a hot plasma of thermalized SM particles. 
To describe the interaction between the Majorana neutrino and the SM particles we adopt the model \eqref{eq3}.  
We further assume that the neutrino mass and the temperature of the plasma satisfy the condition \eqref{scales}.
Symmetry and power counting arguments restrict the form of the corrections and simplify their calculation.
Our result, given in \eqref{eq20}, agrees with earlier findings~\cite{Salvio:2011sf,Laine:2011pq}.
At our accuracy, i.e. first order in the SM couplings and order $T^4/M^3$, 
the two-loop thermal field theory computation necessary to describe the process 
in the full theory splits into two one-loop computations in the EFT.
The first one-loop computation is required to match the full theory with the EFT.
This can be done setting the temperature to zero, so it amounts at the calculation 
of typical in-vacuum matrix elements. The second one-loop computation 
is required to calculate the thermal corrections in the EFT.
At the accuracy of this work, only tadpole diagrams are involved.
These may be easily computed with the real-time formalism or with other methods. 
The use of the real-time formalism is particularly convenient with heavy particles: 
since they do not thermalize, heavy particles and particles coupled to them 
are not affected by the doubling of degrees of freedom typical of the formalism. 
The situation is again analogous to the one faced when studying heavy quarks in a thermal bath.

The total width of the Majorana neutrino, $\Gamma = \Gamma_{T=0} + \Gamma_T$, 
is organized as a double expansion in the SM couplings and in $T/M$.
At the present accuracy, the double expansion reflects the hierarchy of energy scales \eqref{scales}
and corresponds in the EFT to the two steps of the computation: matching and thermal loops. 
The SM couplings entering in the Wilson coefficients of the EFT 
are computed at the heavy neutrino mass scale, $M$.
Whether terms in one expansion are more relevant than terms in the other 
depends on the considered temperature regime. A temperature close to the 
Majorana neutrino mass makes terms in the $T/M$ expansion more relevant, 
although a temperature too close to it may spoil the convergence and 
signal a breakdown of the non-relativistic treatment.

In the future, non-relativistic EFT approaches to heavy Majorana neutrinos 
can be used to simplify computations of the decay rate taking into account
CP violation and a medium away from thermal equilibrium, as well
as in studies of thermal effects within different models.

\acknowledgments

We thank Marco Drewes for discussions. S.B. thanks Alejandro Ibarra 
for comments and suggestions. We acknowledge financial support from the DFG cluster of excellence
\emph{Origin and structure of the universe} (www.universe-cluster.de). 
This research is supported by the DFG grant BR 4058/1-1.

\appendix

\section{Matching and Wilson coefficients}
\label{appA}
In this appendix, we compute the Wilson coefficients \eqref{coa}-\eqref{cod34}.
They are obtained by matching matrix elements calculated in the fundamental theory 
(\ref{eq3}) with matrix elements calculated in the EFT \eqref{eq10}.
The fundamental theory contains the SM with unbroken gauge symmetries, whose Lagrangian reads
\begin{eqnarray}
\mathcal{L}_{\hbox{\tiny {SM}}} &=&
\bar{L}_{f} P_R\, i \slashed{D} \, L_{f} + \bar{Q}P_R\, i\slashed{D} \, Q -\frac{1}{4}W_{\mu\nu}^aW^{a\,\mu\nu} -\frac{1}{4}F_{\mu\nu}F^{\mu\nu}
\nonumber\\
&& + \left( D_{\mu} \phi \right)^{\dagger}\left( D^{\mu} \phi \right)  - \lambda \left( \phi^{\dagger}\phi \right)^2 
- \lambda_{t}    \, \bar{Q} \, \tilde{\phi} \, P_{R} t
- \lambda^{*}_{t} \, \bar{t} P_{L} \, \tilde{\phi}^{\dagger} \, Q  + \dots \,.
\label{SMlag}
\end{eqnarray}
The dots stand for terms that are irrelevant for our calculation, e.g. those involving light quarks or right-handed leptons. 
The covariant derivative is given by 
\begin{equation}
D_{\mu}=\partial_{\mu} -igA^{a}_{\mu} \tau^{a} -ig'YB_{\mu} \, ,
\end{equation}
where $\tau^{a}$ are the SU(2) generators and $Y$ is the hypercharge ($Y=1/2$ for the Higgs, 
$Y=-1/2$ for left-handed leptons). The fields $L_{f}$ are the SU(2) lepton doublets with flavor $f$, 
$Q^T=(t,b)$ is the heavy-quark SU(2) doublet, $t$ is the top quark field, $\phi$ the Higgs doublet, 
$A^{a}_{\mu}$ are the SU(2) gauge fields, $B_{\mu}$ the U(1) gauge fields and 
$W^{a\,\mu\nu}$, $F_{\mu\nu}$ the corresponding field strength tensors.  
The couplings $g$, $g'$, $\lambda$ and $\lambda_t$ are the SU(2) and U(1) gauge 
couplings, the four-Higgs coupling and the top Yukawa coupling respectively.

The effective theory must reproduce the fundamental one at energies below its cut-off~$\Lambda$. 
A way to enforce this is by matching low-energy matrix elements in the two theories.
The matching fixes the Wilson coefficients of the EFT, which encode, order by order in the 
couplings, the contributions from the high-energy modes that have been integrated out.
Because in the matching we are integrating out only high-energy modes, we can set to zero 
any low-energy scale appearing in loops.
In particular, as discussed in the main body of the paper, we can set to zero the temperature.
A consequence is that, in the matching, loop diagrams in the EFT vanish in dimensional 
regularization because scaleless. We adopt dimensional regularization in all loop calculations of the paper.
The Wilson coefficients that we need to compute are those appearing in \eqref{Wa} and \eqref{Wb}.
We compute them by matching four-field matrix elements involving 
two Majorana fields and either two Higgs, two lepton, two quark or two gauge 
fields. We will discuss the matching of these matrix elements one by one in the 
rest of the appendix. Before, we add few general considerations.

We perform the matching in the reference frame $v^{\mu}=(1,\vec{0}\,)$, 
where we assume the plasma to be at rest.
The leading momentum dependent operator \eqref{Wmomdep} is fixed by symmetry and does not need to be calculated. 
Since we are interested in the imaginary parts of the Wilson coefficients,
we evaluate the imaginary parts of $-i {\mathcal{D}}$, 
where ${\mathcal{D}}$ are generic Feynman diagrams,  
by taking the Majorana neutrino mass at $M+i\epsilon$.
We may also choose the incoming and outgoing SM particles to carry the same momentum $q^\mu$.
Because $q^{\mu}$ is much smaller than $M$, diagrams in the fundamental theory are expanded 
in powers of $q^{\mu}$. This expansion matches the operator expansion in the EFT.

The fundamental theory \eqref{eq3} is SU(2)$\times$U(1) gauge invariant, 
so are all operators in the EFT. Hence, the Wilson coefficients are gauge independent.
As a practical choice, however, we will present results for single diagrams in Landau gauge.
This is a convenient gauge in the presence of momentum dependent vertices like those between the Higgs and 
the gauge bosons. We have explicitly checked gauge invariance by computing the Wilson coefficients 
also in Feynman gauge. 

\begin{figure}[htb]
\centering
\includegraphics[scale=0.55]{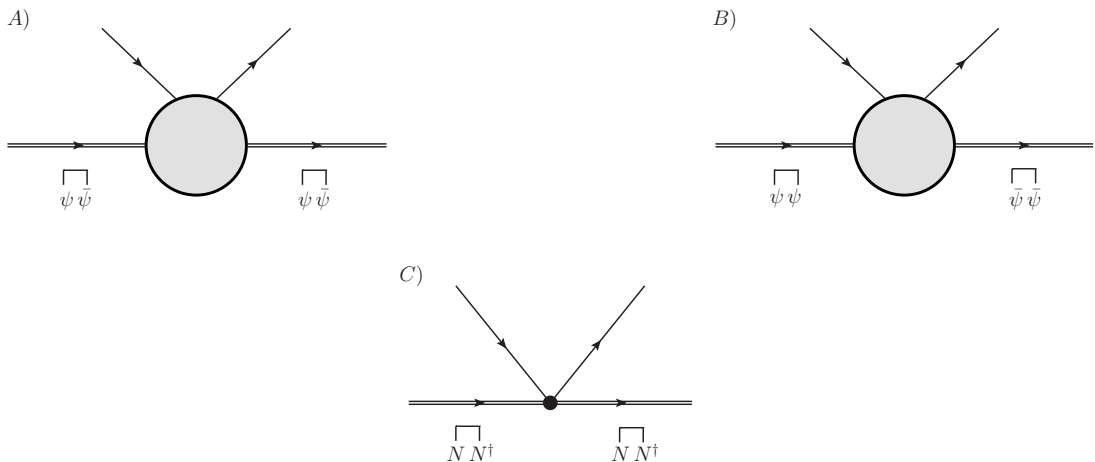}
\caption{\label{Fig0A} The diagrams represent matrix elements with two 
Majorana neutrino fields and two SM fields in the fundamental theory 
(diagrams $A$ and $B$) and in the EFT (diagram $C$). The
bubbles in $A$ and $B$ denote generic loops. 
The diagrams $A$ and $B$ in the relativistic theory
allow for two possible contractions of the neutrino fields, 
while the diagram $C$ in the non-relativistic EFT allows just for one.}
\end{figure}

When computing matrix elements involving Majorana fermions, one has to keep in 
mind that the relativistic Majorana field $\psi$ may be contracted 
in two possible ways, \eqref{eq5} and \eqref{eq6},
as a consequence of the indistinguishability of the particle from the antiparticle. 
A similar observation holds for the field $\bar{\psi}$.
For our calculation, involving matrix elements with two external Majorana neutrinos, 
this implies that in the fundamental theory we have to consider for each diagram 
two possible configurations: each one corresponding to the two possible 
way to contract the Majorana fields $\psi$ and $\bar{\psi}$. See diagrams $A$ and $B$ in figure~\ref{Fig0A}. 
In the non-relativistic EFT, we have only one possible way to contract the Majorana field $N$, 
which is \eqref{effpropagator}. See diagram $C$ in figure~\ref{Fig0A}. 
One has to properly account for this when matching the relativistic matrix elements  
with the ones in the EFT. In our calculation, with the exception of diagrams with external leptons,  
the two possible configurations give the same result as a consequence of 
\begin{equation}
C \gamma^{\mu_1\,T}...\gamma^{\mu_{2n+1}\,T}C=\gamma^{\mu_1}...\gamma^{\mu_{2n+1}} \, ,
\end{equation}  
and because of the insensitivity of the result to the direction of the momentum carried by the Majorana neutrino.

\subsection{Higgs}
\begin{figure}[htb]
\centering
\includegraphics[scale=0.435]{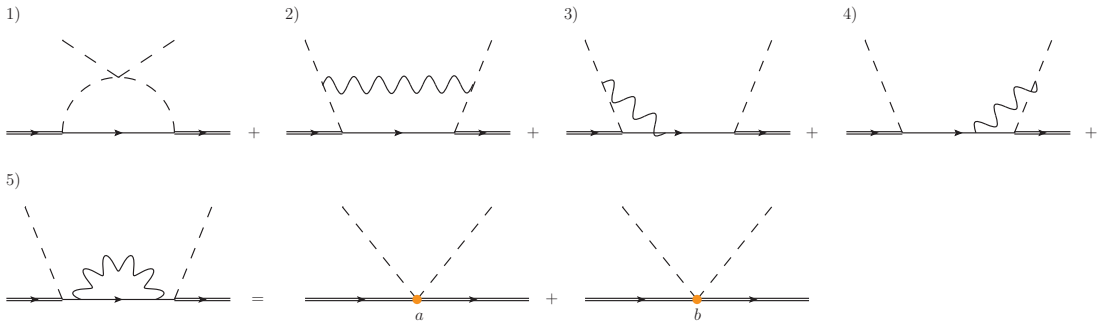}
\caption{\label{Fig1A} Diagrams in the full theory (left-hand side of the equality)
contributing to the Majorana neutrino-Higgs four-field operators in the EFT (right-hand side). 
The solid double lines stand for heavy Majorana neutrinos, the solid single lines 
for leptons, the dashed lines for Higgs particles and the wiggled lines for gauge bosons.}
\end{figure}

In order to determine the Wilson coefficients $a$ and $b$, we compute in the fundamental theory the matrix element
\begin{equation}
-i \left.\int d^{4}x\,e^{i p \cdot x} \int d^{4}y \int d^{4}z\,e^{i q \cdot (y-z)}\, 
\langle \Omega | T(\psi^{\mu}(x) \bar{\psi}^{\nu }(0) \phi_{m}(y) \phi_{n}^{\dagger}(z) )| \Omega \rangle
\right|_{p^\mu =(M + i\epsilon,\vec{0}\,)},
\label{A1}
\end{equation} 
where $\mu$ and $\nu$ are Lorentz indices, $m$ and $n$ are SU(2) indices and $|\Omega \rangle$ is the ground state 
of the fundamental theory. The matrix element \eqref{A1} describes a $2 \rightarrow 2$ scattering between 
a heavy Majorana neutrino at rest and a Higgs boson carrying momentum $q^\mu$.
In figure~\ref{Fig1A}, we show on the left-hand side of the equality all diagrams 
that in the fundamental theory contribute to the effective vertices shown on its right-hand side. 

In order to compute the imaginary parts of the Wilson coefficients $a$ and $b$, 
we need to consider only the imaginary parts of the diagrams shown in figure~\ref{Fig1A}.
In Landau gauge, the diagrams in the fundamental theory read\footnote{ 
To keep the notation simple, we drop, from now and in the rest of the appendix, 
propagators on external legs, and we label the so-obtained amputated Green's functions 
with the same indices used for the unamputated ones. 
}
\begin{eqnarray}
&& {\rm{Im}}\, (-i \mathcal{D}_{1}) = 
- \frac{3}{8 \pi} \frac{\lambda|F|^{2}}{M}\delta_{mn} \delta^{\mu \nu} + \dots \, ,   
\\
&& {\rm{Im}}\, (-i \mathcal{D}_{2}) =  
- \frac{1}{96 \pi} \frac{(3g^{2}+g'^{\,2})|F|^{2}}{M^{3}}\delta_{mn}\delta^{\mu \nu}    (q_{0})^{2}  + \dots \, ,  
\\
&& {\rm{Im}}\,(-i \mathcal{D}_{3}) + {\rm{Im}}\,(-i \mathcal{D}_{4}) = 
- \frac{7}{48 \pi} \frac{(3g^{2}+g'^{\,2})|F|^{2}}{M^ {3}}\delta_{mn}\delta^{\mu \nu}   (q_{0})^{2} + \dots \, ,   
\\
&& {\rm{Im}}\,(-i \mathcal{D}_5) = 0\, ,    
\end{eqnarray}
where the subscripts refer to the diagrams as listed in figure~\ref{Fig1A}.\footnote{
The vanishing of diagram 5 is specific of the Landau gauge.}
The dots stand for terms that are either proportional to $q^\mu/M^2$, or to 
$q_0q_i/M^3$ ($i=1,2,3$) or to $q^2/M^3$; we have not displayed terms that are of order $1/M^4$ or smaller.
Such terms do not contribute to the matching of the operators in \eqref{Wa} and \eqref{Wb}. 
Summing up all contributions we get 
\begin{equation}
- \frac{3}{8 \pi} \frac{\lambda|F|^{2}}{M}\delta_{mn}\delta^{\mu \nu}  
- \frac{5}{32 \pi} \frac{(3g^{2}+g'^{\,2})|F|^{2}}{M^ {3}}\delta_{mn} \delta^{\mu \nu}  (q_{0})^{2}  + \dots \, .
\label{A10}
\end{equation}  

The symmetries of the EFT enforce that the matrix element \eqref{A1} is reproduced by the following expression 
\begin{equation}
\frac{a}{M} \delta_{mn}\delta^{\mu \nu}  +   \frac{b}{M^{3}} \delta_{mn}\delta^{\mu \nu} (q_{0})^{2} + \dots \,,
\label{A3}
\end{equation}
where the dots stand for contributions coming from operators that are not listed in \eqref{Wa} and \eqref{Wb}. 

Matching the imaginary part of (\ref{A3}) with (\ref{A10}) fixes the imaginary parts of $a$ and $b$:
\begin{equation}
{\rm Im}\,a = -\frac{3}{8\pi}|F|^{2}\lambda \, , 
\qquad 
{\rm Im}\,b = - \frac{5}{32 \pi} (3g^{2}+g'^{\,2})|F|^{2} \, .
\end{equation} 
Note that only the first diagram of figure~\ref{Fig1A}
contributes to the effective operator \eqref{Wa}, which provides the leading contribution 
to the Majorana neutrino thermal width. The remaining diagrams contribute to the subleading 
operator $b \; N^{\dagger}N \, (D_0 \phi^{\dagger}) \, (D_0 \phi)/M^3$.

\begin{figure}[htb]
\includegraphics[scale=0.384]{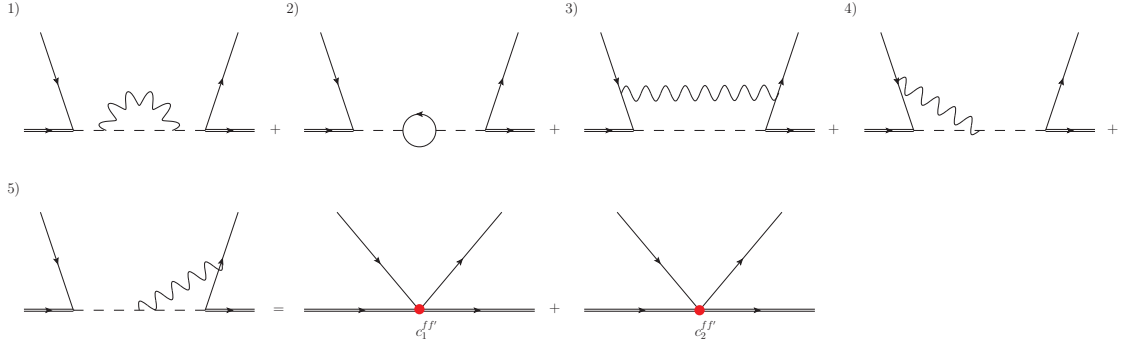}
\caption{\label{Fig2A} 
Diagrams in the full theory (left-hand side of the equality)
contributing to the Majorana neutrino-lepton four-fermion operators in the EFT (right-hand side).
The lines stand for the same particle propagators as in figure~\ref{Fig1A}.
}
\end{figure}

\subsection{Leptons}
In the fundamental theory, the matrix element 
\begin{equation}
-i \left.\int d^{4}x\,e^{i p \cdot x} \int d^{4}y \int d^{4}z\,e^{i q \cdot (y-z)}\, 
\langle \Omega | T(\psi^{\mu}(x) \bar{L}^{\beta}_{f,m}(z) L^{\alpha}_{f',n}(y) \bar{\psi}^{\nu }(0))  | \Omega \rangle 
\right|_{p^\mu =(M + i\epsilon,\vec{0}\,)},
\label{B1}
\end{equation}
where $f$ and $f'$ are flavor indices, $\alpha$, $\beta$, $\mu$ and $\nu$ Lorentz indices, and $m$ and $n$ SU(2) indices, 
describes a $2 \rightarrow 2$ scattering between a heavy Majorana neutrino at rest and a lepton carrying momentum $q^\mu$.
The diagrams contributing to the matrix element in the fundamental theory are shown on the left-hand side of the equality 
of figure~\ref{Fig2A}. 
Their imaginary part in Landau gauge gives 
\begin{eqnarray}
{\rm{Im}}\,(-i\mathcal{D}_{1}) &=& 
- \delta_{mn} F_{f'} F^{*}_{f} \left( \frac{3(3g^{2}+g'^{\,2})}{32 \pi M^{3}}\right) 
\left[(P_{L})^{\mu \beta}(P_{R})^{ \alpha \nu} \right.
\nonumber\\
&& \hspace{4.0cm} \left.
+ (C\,P_{R})^{\mu \alpha}(P_{L}\,C)^{\beta \nu}\right] q_{0} 
+\dots\,,
\\
{\rm{Im}}\,(-i\mathcal{D}_{2}) &=&  
\delta_{mn} F_{f'} F^{*}_{f} \left( \frac{3|\lambda_{t}|^2}{8 \pi M^{3}}\right)     
\left[(P_{L})^{\mu \beta}(P_{R})^{ \alpha \nu} \right.
\nonumber\\
&& \hspace{4.0cm} \left.
+ (C\,P_{R})^{\mu \alpha}(P_{L}\,C)^{\beta \nu}\right] q_{0}
+\dots\,,
\\
{\rm{Im}}\,(-i\mathcal{D}_{3}) &=& 
- \delta_{mn} F_{f'} F^{*}_{f} \left( \frac{(3g^{2}+g'^{\,2})}{32 \pi M^{3}} \right) 
\left[(P_{L})^{\mu \beta}(P_{R})^{ \alpha \nu} + (C\,P_{R})^{\mu \alpha}(P_{L}\,C)^{\beta \nu}\right] q_{0}
\nonumber \\
&&+ \delta_{mn} F_{f'} F^{*}_{f}  \left( \frac{(3g^{2}+g'^{\,2})}{384 \pi M^{3}} \right)    
\left[ (P_{L}\,\gamma_{\lambda}\gamma_{\sigma})^{\mu \beta}(\gamma^{\sigma}\gamma^{\lambda}\, P_{R})^{ \alpha \nu} \right.
\nonumber\\
&& \hspace{4.0cm} \left.
+ (C\, P_{R}\,\gamma_{\lambda}\gamma_{\sigma})^{\mu \alpha}(\gamma^{\sigma}\gamma^{\lambda} \,P_{L}\,C)^{\beta\nu} \right]q_{0}  
+\dots\,,
\\
{\rm{Im}}\,(-i\mathcal{D}_{4}) &+& {\rm{Im}}\,(-i\mathcal{D}_{5}) = 
- \delta_{mn} F_{f'} F^{*}_{f} \left( \frac{(3g^{2}+g'^{\,2})}{16 \pi M^{3}}\right) 
\left[(P_{L})^{\mu \beta}(P_{R})^{ \alpha \nu} \right.
\nonumber\\
&& \hspace{4.0cm} \left.
+ (C\,P_{R})^{\mu \alpha}(P_{L}\,C)^{\beta \nu}\right] q_{0} 
+\dots\,.
\label{B2}
\end{eqnarray}
where the subscripts refer to the diagrams as listed in figure~\ref{Fig2A} and the dots stand either for higher-order terms 
in the $1/M$ expansion or for terms of order $1/M^2$ but that do not depend on the momentum $q^\mu$.
Summing up all contributions and comparing with the corresponding expression in the EFT, which is 
\begin{eqnarray}
&& \hspace{-1cm}
\frac{c^{ff'}_{1}}{M^{3}} \delta_{mn} \left[(P_{L})^{\mu \beta}(P_{R})^{ \alpha \nu} + (C\,P_{R})^{\mu \alpha}(P_{L}\,C)^{\beta \nu}\right] q_{0} 
\nonumber\\
&& \hspace{-1cm}
+ \frac{c^{ff'}_{2}}{M^{3}} \delta_{mn} 
\left[ (P_{L}\,\gamma_{\lambda}\gamma_{\sigma})^{\mu \beta}(\gamma^{\sigma}\gamma^{\lambda} \,P_{R})^{ \alpha \nu} 
+ (C\, P_{R}\,\gamma_{\lambda}\gamma_{\sigma})^{\mu \alpha}(\gamma^{\sigma}\gamma^{\lambda} \,P_{L}\,C)^{\beta\nu} \right] q_{0} 
+\dots\,,
\label{B4}
\end{eqnarray}
we obtain
\begin{equation}
{\rm Im}\,c^{ff'}_{1} = \frac{3}{8\pi}|\lambda_{t}|^{2}F_{f'} F^{*}_{f}  - \frac{3}{16 \pi}(3g^{2}+g'^{\,2}) F_{f'} F^{*}_{f} \,, 
\qquad {\rm Im}\,c^{ff'}_{2} =  \frac{1}{384 \pi} (3g^{2}+g'^{\,2})F_{f'} F^{*}_{f} \, .
\label{B5}
\end{equation} 
The dots in \eqref{B4} stand for contributions coming from operators that are not listed in~\eqref{Wb}.

\begin{figure}[htb]
\centering
\includegraphics[scale=0.4]{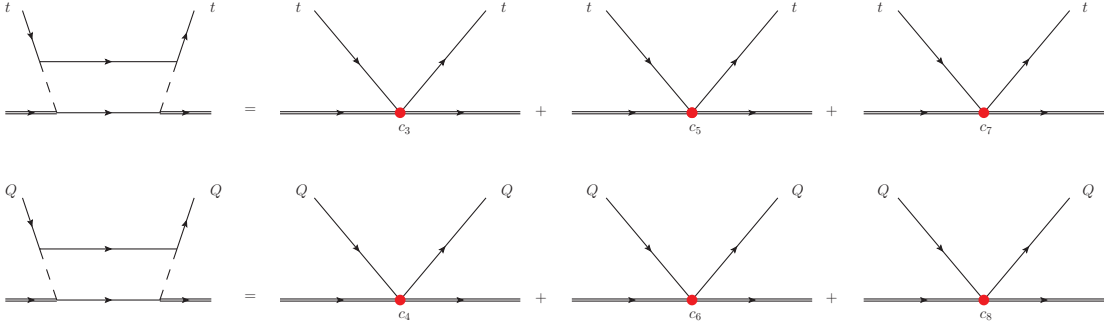}
\caption{\label{Fig3A} In the top panel, the diagram in the full theory (left-hand side)
contributing to the Majorana neutrino-top-quark singlet four-fermion operators in the EFT (right-hand side).
In the bottom panel, the diagram in the full theory (left-hand side)
contributing to the Majorana neutrino-heavy-quark doublet four-fermion operators in the EFT (right-hand side).
The solid single lines marked $t$ stand for top singlets, the solid single lines marked $Q$ 
for heavy-quark doublets, unmarked  solid lines connecting top lines and heavy-quark doublets stand for 
heavy-quark doublets and top singlets respectively. All other lines stand for the same particle propagators as in figure~\ref{Fig1A}.}
\end{figure}

\subsection{Quarks}
We consider only couplings with top quarks, for $\lambda_{t} \sim 1$ while 
all other Yukawa couplings are negligible.
In the fundamental theory, we compute the two matrix elements
\begin{eqnarray}
&&\hspace{-10mm}
-i \left.\int d^{4}x\,e^{i p \cdot x} \int d^{4}y \int d^{4}z\,e^{i q \cdot (y-z)}\, 
\langle \Omega | T(\psi^{\mu}(x) \bar{\psi}^{\nu }(0) \, t^{\alpha}(y) \bar{t}^{\beta}(z)) | \Omega \rangle 
\right|_{p^\mu =(M + i\epsilon,\vec{0}\,)}\!,
\label{C11}\\
&&\hspace{-10mm}
-i \left.\int d^{4}x\,e^{i p \cdot x} \int d^{4}y \int d^{4}z\,e^{i q \cdot (y-z)}\, 
\langle \Omega | T(\psi^{\mu}(x) \bar{\psi}^{\nu }(0) \, Q_{m}^{\alpha}(y) \bar{Q}_{n}^{\beta}(z)) | \Omega \rangle 
\right|_{p^\mu =(M + i\epsilon,\vec{0}\,)}\!,
\label{C1}
\end{eqnarray}
describing respectively a $2 \rightarrow 2$ scattering between a heavy Majorana neutrino at rest and a right-handed top quark carrying momentum $q^\mu$, 
and a $2 \rightarrow 2$ scattering between a heavy Majorana neutrino at rest and a left-handed heavy quark carrying momentum $q^\mu$. 
The indices $\alpha$, $\beta$, $\mu$ and $\nu$ are Lorentz indices, whereas $m$ and $n$ are the SU(2) indices 
of the heavy-quark doublet.
The diagrams contributing to the matrix elements in the fundamental theory are shown in figure~\ref{Fig3A}. 
We call $\mathcal{D}_{t}$ the diagram with external top lines and  $\mathcal{D}_{Q}$ the diagram with external 
heavy-quark lines. The imaginary parts of $-i\mathcal{D}_{t}$ and $-i\mathcal{D}_{Q}$ read 
\begin{eqnarray}
&&\hspace{-10mm}
{\rm{Im}}\,(-i\mathcal{D}_{t}) = 
\frac{|F|^{2}|\lambda_{t}|^2}{24 \pi M^{3}}   \delta^{\mu \nu}   \left( P_{L} \gamma^{0} \right)^{\alpha \beta} q_{0} 
\nonumber\\
&& \hspace{10mm}
+ \frac{|F|^{2}|\lambda_{t}|^2}{48 \pi M^{3}}\left[   
\left(\gamma^5\gamma^i\right)^{\mu \nu}   \left( P_{L} \gamma^{0} \right)^{\alpha \beta} q_{i}
+\left(\gamma^5\gamma^i\right)^{\mu \nu}  \left( P_{L} \gamma_{i} \right)^{\alpha \beta} q_{0}
\right] + \dots\,,
\label{C2} \\
&&\hspace{-10mm}
{\rm{Im}}\,(-i\mathcal{D}_{Q}) = 
\frac{|F|^{2}|\lambda_{t}|^2}{48 \pi M^{3}}  \delta_{mn} \delta^{\mu \nu} \left( P_{R} \gamma^{0} \right)^{\alpha \beta}   q_{0} 
\nonumber\\
&& \hspace{10mm}
+ \frac{|F|^{2}|\lambda_{t}|^2}{96 \pi M^{3}} \delta_{mn} \left[   
\left(\gamma^5\gamma^i\right)^{\mu \nu}   \left( P_{R} \gamma^{0} \right)^{\alpha \beta} q_{i}
+\left(\gamma^5\gamma^i\right)^{\mu \nu}  \left( P_{R} \gamma_{i} \right)^{\alpha \beta} q_{0}
\right] + \dots\,,
\label{C3}
\end{eqnarray}
where the dots stand for higher-order terms in the $1/M$ expansion or terms that are of order $1/M^2$ 
but do not depend on the momentum $q^\mu$. 

The matrix element \eqref{C11} is matched in the EFT by 
\begin{equation}
\frac{c_{3}}{M^{3}}  \delta^{\mu \nu}   \left( P_{L} \gamma^{0} \right)^{\alpha \beta}  q_{0}   
+ \frac{c_{5}}{M^{3}}  \left(\gamma^5\gamma^i\right)^{\mu \nu}  \left( P_{L} \gamma^{0} \right)^{\alpha \beta} q_{i}
+ \frac{c_{7}}{M^{3}}  \left(\gamma^5\gamma^i\right)^{\mu \nu}  \left( P_{L} \gamma_{i} \right)^{\alpha \beta} q_{0}
+ \dots,
\label{C6}
\end{equation}
and the matrix element \eqref{C1} by
\begin{equation}
\frac{c_{4}}{M^{3}}  \delta_{mn} \delta^{\mu \nu}  \left( P_{R} \gamma^{0} \right)^{\alpha \beta}      q_{0} 
+ \frac{c_{6}}{M^{3}} \delta_{mn}\left(\gamma^5\gamma^i\right)^{\mu \nu}  \left( P_{R} \gamma^{0} \right)^{\alpha \beta} q_{i}
+ \frac{c_{8}}{M^{3}} \delta_{mn}\left(\gamma^5\gamma^i\right)^{\mu \nu}  \left( P_{R} \gamma_{i} \right)^{\alpha \beta} q_{0}
+ \dots,
\label{C7}
\end{equation}
where the dots in \eqref{C6} and \eqref{C7} stand for contributions coming from operators not listed in~\eqref{Wb}.
Comparing \eqref{C2} and \eqref{C3} with the imaginary parts of \eqref{C6} and \eqref{C7} respectively, we obtain 
\begin{eqnarray}
{\rm Im}\,c_{3}= \frac{1}{24\pi}|\lambda_{t}|^{2}|F|^{2} \,, && \qquad {\rm Im}\,c_{4}= \frac{1}{48\pi}|\lambda_{t}|^{2}|F|^{2} \,,
\label{C81}\\
{\rm Im}\,c_{5}= \frac{1}{48\pi}|\lambda_{t}|^{2}|F|^{2} \,, && \qquad {\rm Im}\,c_{6}= \frac{1}{96\pi}|\lambda_{t}|^{2}|F|^{2} \,,
\label{C82}\\
{\rm Im}\,c_{7}= \frac{1}{48\pi}|\lambda_{t}|^{2}|F|^{2} \,, && \qquad {\rm Im}\,c_{8}= \frac{1}{96\pi}|\lambda_{t}|^{2}|F|^{2} \,.
\label{C83}
\end{eqnarray}

\begin{figure}[htb]
\centering
\includegraphics[scale=0.46]{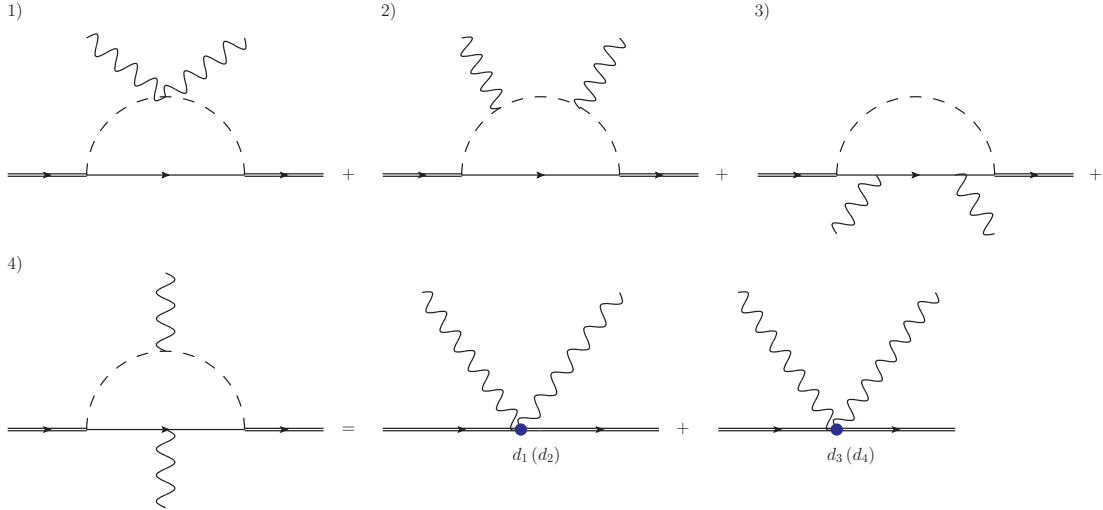}
\caption{\label{Fig4A} 
Diagrams in the full theory (left-hand side of the equality)
contributing to the Majorana neutrino-gauge boson four-field operators in the EFT (right-hand side).
Diagrams with crossed gauge bosons have not been explicitly displayed.
External gauge fi\-elds are ei\-ther SU(2) or U(1) gauge fields. 
In one case they contribute to the operators $d_{1} \, N^{\dagger}N \, W^{a}_{i0} W^a_{i0}/M^3$ 
and $d_{3} \, N^\dagger N \, W^{a}_{\mu\nu} W^{a\,\mu\nu}\!/M^3$, 
in the other case to the operators $d_{2} \, N^{\dagger}N \, F_{i0} F_{i0}/M^3$ 
and $d_{4} \, N^\dagger N $ $\times F_{\mu\nu} F^{\mu\nu}\!/M^3$ in the EFT.
The lines stand for the same particle propagators as in figure~\ref{Fig1A}.
}
\end{figure}

\subsection{Gauge bosons}
The couplings $d_{i}$ of the Majorana neutrino with the gauge bosons 
are conveniently computed by considering in the fundamental theory the following two matrix elements 
\begin{equation}
-i \left.\int d^{4}x\,e^{i p \cdot x} \int d^{4}y \int d^{4}z\,e^{i q \cdot (y-z)}\, 
\langle \Omega | T(\psi^{\mu}(x) \bar{\psi}^{\nu}(0)  \, A^a_i(y) \, A^b_j(z) )| \Omega\rangle 
\right|_{p^\mu =(M + i\epsilon,\vec{0}\,)},
\label{D10}
\end{equation}
and 
\begin{equation}
-i \left.\int d^{4}x\,e^{i p \cdot x} \int d^{4}y \int d^{4}z\,e^{i q \cdot (y-z)}\, 
\langle \Omega | T(\psi^{\mu}(x) \bar{\psi}^{\nu}(0)  \, B_i(y) \, B_j(z) )| \Omega\rangle 
\right|_{p^\mu =(M + i\epsilon,\vec{0}\,)},
\label{D1}
\end{equation}
where $a$ and $b$ are indices labeling fields in the adjoint representation of SU(2), and $i$ and $j$ are spatial Lorentz indices.
The matrix elements \eqref{D10} and \eqref{D1} describe $2 \rightarrow 2$ scatterings between heavy Majorana neutrinos 
at rest and gauge bosons carrying momentum $q^\mu$. 
Each diagram in the full theory, labeled according to figure~\ref{Fig4A},  
contributes with an imaginary part that reads for the \eqref{D10} matrix element 
\begin{eqnarray}
{\rm{Im}}\,(-i\mathcal{D}_{1}) &=& 
- \frac{g^{2} |F|^{2}}{16 \pi M} \,  \delta^{\mu \nu} \delta^{ab}  \delta_{ij} 
+\dots \,,   
\label{D41}\\
{\rm{Im}}\,(-i\mathcal{D}_{2}) &=& 
\frac{g^{2} |F|^{2}}{16 \pi M}  \, \delta^{\mu \nu}  \delta^{ab} 
\left( \delta_{ij} + \delta_{ij} \frac{(q_{0})^{2}}{3 M^{2}}  + \frac{q_iq_j}{6 M^{2}} \right)   
+\dots \,,   
\label{D42}\\
{\rm{Im}}\,(-i\mathcal{D}_{3}) &=& 
- \frac{g^{2} |F|^{2}}{24 \pi M^{3}}  \, \delta^{\mu \nu} \delta^{ab}  \, 
\left( \delta_{ij} (q_{0})^{2} - \frac{q_iq_j}{2} \right) 
+\dots \,,   
\label{D43}\\
{\rm{Im}}\,(-i\mathcal{D}_{4}) &=& 
- \frac{g^{2} |F|^{2}}{48 \pi M^{3}}  \, \delta^{\mu \nu} \delta^{ab}  \, q_iq_j 
+\dots \,.
\label{D44}
\end{eqnarray}
For the matrix element \eqref{D1} the result is the same after the replacement 
$g^2\delta^{ab} \to g'^2$. The dots stand for $1/M^3$ terms that are proportional to $q^2$ or $q_0q_i$ 
or for terms of order $1/M^4$ or smaller.

The matrix element \eqref{D10} is matched in the EFT by 
\begin{equation}
\frac{2 d_{1} }{M^{3}} \, \delta^{\mu \nu}  \delta^{ab} \delta_{ij} \, (q_{0})^{2}  
- \frac{4 d_{3} }{M^{3}} \, \delta^{\mu \nu}  \delta^{ab} \, q_iq_j  
+\dots\,,
\label{D2}
\end{equation}
and the matrix element \eqref{D1} by 
\begin{equation}
\frac{2 d_{2} }{M^{3}} \, \delta^{\mu \nu}  \delta_{ij} \, (q_{0})^{2}  
- \frac{4 d_{4} }{M^{3}} \, \delta^{\mu \nu}  \, q_iq_j  
+\dots\,,
\label{D3}
\end{equation}
where the dots stand for contributions coming from operators not listed in~\eqref{Wb}.
Summing up all contributions \eqref{D41}-\eqref{D44} for each of the two matrix elements and comparing 
with the imaginary parts of \eqref{D2} and \eqref{D3}, we finally find
\begin{eqnarray}
{\rm{Im}}\,d_{1}= -\frac{g^{2}|F|^{2}}{96 \pi} \,, && \qquad {\rm{Im}}\,d_{2}=-\frac{g'^{\,2}|F|^{2}}{96 \pi} \,,
\label{D5}\\
{\rm{Im}}\,d_{3}= -\frac{g^{2}|F|^{2}}{384 \pi} \,, && \qquad {\rm{Im}}\,d_{4}=-\frac{g'^{\,2}|F|^{2}}{384 \pi} \,.
\label{D51}
\end{eqnarray}
The same Wilson coefficients satisfy the matching conditions for matrix elements with temporal gauge bosons.


\begin{thebibliography}{99}

%\cite{Drewes:2013gca}
\bibitem{Drewes:2013gca}
  M.~Drewes,
  %``The Phenomenology of Right Handed Neutrinos,''
  arXiv:1303.6912 [hep-ph].
  %%CITATION = ARXIV:1303.6912;%%

%\cite{Fukuda:1998mi}
\bibitem{Fukuda:1998mi}
  Y.~Fukuda {\it et al.}  [Super-Kamiokande Collaboration],
  %``Evidence for oscillation of atmospheric neutrinos,''
  Phys.\ Rev.\ Lett.\  {\bf 81} (1998) 1562
  [hep-ex/9807003].
  %%CITATION = HEP-EX/9807003;%%

%\cite{Ahmed:2003kj}
\bibitem{Ahmed:2003kj}
  S.~N.~Ahmed {\it et al.}  [SNO Collaboration],
  %``Measurement of the total active B-8 solar neutrino flux at the Sudbury Neutrino Observatory with enhanced neutral current sensitivity,''
  Phys.\ Rev.\ Lett.\  {\bf 92} (2004) 181301
  [nucl-ex/0309004].
  %%CITATION = NUCL-EX/0309004;%%

%\cite{Minkowski:1977sc}
\bibitem{Minkowski:1977sc}
  P.~Minkowski,
  %``mu --> e gamma at a Rate of One Out of 1-Billion Muon Decays?,''
  Phys.\ Lett.\ B {\bf 67} (1977) 421.
  %%CITATION = PHLTA,B67,421;%%

%\cite{GellMann:1980vs}
\bibitem{GellMann:1980vs}
  M.~Gell-Mann, P.~Ramond and R.~Slansky,
  %``Complex Spinors And Unified Theories,''
  Conf.\ Proc.\ C {\bf 790927} (1979) 315
  [arXiv:1306.4669 [hep-th]].
  %%CITATION = ARXIV:1306.4669;%%

%\cite{Mohapatra:1979ia}
\bibitem{Mohapatra:1979ia}
  R.~N.~Mohapatra and G.~Senjanovic,
  %``Neutrino Mass and Spontaneous Parity Violation,''
  Phys.\ Rev.\ Lett.\  {\bf 44} (1980) 912.
  %%CITATION = PRLTA,44,912;%%

%\cite{Dolgov:1991fr}
\bibitem{Dolgov:1991fr}
  A.~D.~Dolgov,
  %``NonGUT baryogenesis,''
  Phys.\ Rept.\  {\bf 222} (1992) 309.
  %%CITATION = PRPLC,222,309;%%

%\cite{Komatsu:2008hk}
\bibitem{Komatsu:2008hk}
  E.~Komatsu {\it et al.}  [WMAP Collaboration],
  %``Five-Year Wilkinson Microwave Anisotropy Probe (WMAP) Observations: Cosmological Interpretation,''
  Astrophys.\ J.\ Suppl.\  {\bf 180} (2009) 330
  [arXiv:0803.0547 [astro-ph]].
  %%CITATION = ARXIV:0803.0547;%%

%\cite{Fukugita:1986hr}
\bibitem{Fukugita:1986hr}
  M.~Fukugita and T.~Yanagida,
  %``Baryogenesis Without Grand Unification,''
  Phys.\ Lett.\ B {\bf 174} (1986) 45.
  %%CITATION = PHLTA,B174,45;%%

%\cite{Bertone:2004pz}
\bibitem{Bertone:2004pz}
  G.~Bertone, D.~Hooper and J.~Silk,
  %``Particle dark matter: Evidence, candidates and constraints,''
  Phys.\ Rept.\  {\bf 405} (2005) 279
  [hep-ph/0404175].
  %%CITATION = HEP-PH/0404175;%%

%\cite{Boyarsky:2009ix}
\bibitem{Boyarsky:2009ix}
  A.~Boyarsky, O.~Ruchayskiy and M.~Shaposhnikov,
  %``The Role of sterile neutrinos in cosmology and astrophysics,''
  Ann.\ Rev.\ Nucl.\ Part.\ Sci.\  {\bf 59} (2009) 191
  [arXiv:0901.0011 [hep-ph]].
  %%CITATION = ARXIV:0901.0011;%%

%\cite{Isgur:1989vq}
\bibitem{Isgur:1989vq}
  N.~Isgur and M.~B.~Wise,
  %``Weak Decays of Heavy Mesons in the Static Quark Approximation,''
  Phys.\ Lett.\ B {\bf 232} (1989) 113.
  %%CITATION = PHLTA,B232,113;%%

%\cite{Eichten:1989zv}
\bibitem{Eichten:1989zv}
  E.~Eichten and B.~R.~Hill,
  %``An Effective Field Theory for the Calculation of Matrix Elements Involving Heavy Quarks,''
  Phys.\ Lett.\ B {\bf 234} (1990) 511.
  %%CITATION = PHLTA,B234,511;%%

%\cite{Kopp:2011gg}
\bibitem{Kopp:2011gg}
  K.~Kopp and T.~Okui
  %``Effective Field Theory for a Heavy Majorana Fermion,''
  Phys.\ Rev.\ D {\bf 84} (2011) 093007
  [arXiv:1108.2702 [hep-ph]].
  %%CITATION = ARXIV:1108.2702;%%

%\cite{Salvio:2011sf}
\bibitem{Salvio:2011sf}
  A.~Salvio, P.~Lodone and A.~Strumia,
  %``Towards leptogenesis at NLO: the right-handed neutrino interaction rate,''
  JHEP {\bf 1108} (2011) 116
  [arXiv:1106.2814 [hep-ph]].
  %%CITATION = ARXIV:1106.2814;%%

%\cite{Laine:2011pq}
\bibitem{Laine:2011pq}
  M.~Laine and Y.~Schr\"oder,
  %``Thermal right-handed neutrino production rate in the non-relativistic regime,''
  JHEP {\bf 1202} (2012) 068
  [arXiv:1112.1205 [hep-ph]].
  %%CITATION = ARXIV:1112.1205;%%

%\cite{Neubert:1993mb}
\bibitem{Neubert:1993mb}
  M.~Neubert,
  %``Heavy quark symmetry,''
  Phys.\ Rept.\  {\bf 245} (1994) 259
  [hep-ph/9306320].
  %%CITATION = HEP-PH/9306320;%%

%\cite{Dugan:1991ak}
\bibitem{Dugan:1991ak}
  M.~J.~Dugan, M.~Golden and B.~Grinstein,
  %``On the Hilbert space of the heavy quark effective theory,''
  Phys.\ Lett.\ B {\bf 282} (1992) 142.
  %%CITATION = PHLTA,B282,142;%%

%\cite{Mannheim:1980eb}
\bibitem{Mannheim:1980eb}
  P.~D.~Mannheim,
  %``Theory Of Majorana Masses,''
  Int.\ J.\ Theor.\ Phys.\  {\bf 23} (1984) 643.
  %%CITATION = IJTPB,23,643;%%

%\cite{Hill:2011be}
\bibitem{Hill:2011be}
  R.~J.~Hill and M.~P.~Solon,
  %``Universal behavior in the scattering of heavy, weakly interacting dark matter on nuclear targets,''
  Phys.\ Lett.\ B {\bf 707} (2012) 539
  [arXiv:1111.0016 [hep-ph]].
  %%CITATION = ARXIV:1111.0016;%%

%\cite{Luty:1992un}
\bibitem{Luty:1992un}
  M.~A.~Luty,
  %``Baryogenesis via leptogenesis,''
  Phys.\ Rev.\ D {\bf 45} (1992) 455.
  %%CITATION = PHRVA,D45,455;%%

%\cite{Buchmuller:2005eh}
\bibitem{Buchmuller:2005eh}
  W.~Buchm\"uller, R.~D.~Peccei and T.~Yanagida,
  %``Leptogenesis as the origin of matter,''
  Ann.\ Rev.\ Nucl.\ Part.\ Sci.\  {\bf 55} (2005) 311
  [hep-ph/0502169].
  %%CITATION = HEP-PH/0502169;%%

%\cite{Davidson:2008bu}
\bibitem{Davidson:2008bu}
  S.~Davidson, E.~Nardi and Y.~Nir,
  %``Leptogenesis,''
  Phys.\ Rept.\  {\bf 466} (2008) 105
  [arXiv:0802.2962 [hep-ph]].
  %%CITATION = ARXIV:0802.2962;%%

%\cite{Asaka:2005pn}
\bibitem{Asaka:2005pn}
  T.~Asaka and M.~Shaposhnikov,
  %``The nuMSM, dark matter and baryon asymmetry of the universe,''
  Phys.\ Lett.\ B {\bf 620} (2005) 17
  [hep-ph/0505013].
  %%CITATION = HEP-PH/0505013;%%

%\cite{Asaka:2005an}
\bibitem{Asaka:2005an}
  T.~Asaka, S.~Blanchet and M.~Shaposhnikov,
  %``The nuMSM, dark matter and neutrino masses,''
  Phys.\ Lett.\ B {\bf 631} (2005) 151
  [hep-ph/0503065].
  %%CITATION = HEP-PH/0503065;%%

%\cite{Asaka:2006rw}
\bibitem{Asaka:2006rw}
  T.~Asaka, M.~Laine and M.~Shaposhnikov,
  %``On the hadronic contribution to sterile neutrino production,''
  JHEP {\bf 0606} (2006) 053
  [hep-ph/0605209].
  %%CITATION = HEP-PH/0605209;%%

%\cite{Brambilla:2003nt}
\bibitem{Brambilla:2003nt}
  N.~Brambilla, D.~Gromes and A.~Vairo,
  %``Poincare invariance constraints on NRQCD and potential NRQCD,''
  Phys.\ Lett.\  B {\bf 576} (2003) 314
  [arXiv:hep-ph/0306107].
  %%CITATION = PHLTA,B576,314;%%

%\cite{LeBellac:1996}
\bibitem{LeBellac:1996}
  M.~Le~Bellac, {\it Thermal Field Theory}, Cambridge University Press,  Cambridge U.K. (1996). 
%M. Le Bellac, Thermal Field Theory (Cambridge University Press, Cambridge 1996).
  
%\cite{Brambilla:2008cx}
\bibitem{Brambilla:2008cx}
  N.~Brambilla, J.~Ghiglieri, A.~Vairo and P.~Petreczky,
  %``Static quark-antiquark pairs at finite temperature,''
  Phys.\ Rev.\ D {\bf 78} (2008) 014017
  [arXiv:0804.0993 [hep-ph]].
  %%CITATION = ARXIV:0804.0993;%%

\end{thebibliography}
\end{document}